\documentclass{article}%
\usepackage{amssymb}
\usepackage{amsmath}
\usepackage{amsfonts}
\usepackage{graphicx}%
\providecommand{\U}[1]{\protect\rule{.1in}{.1in}}

\begin{document}

\title{Jorge A. Swieca's contributions to quantum field theory in the 60s and 70s and
their relevance in present research \\{\small to be published in EPJH - Historical Perspectives on Contemporary
Physics}}
\author{Bert Schroer\\CBPF, Rua Dr. Xavier Sigaud 150 \\22290-180 Rio de Janeiro, Brazil\\and Institut f\"{u}r Theoretische Physik der FU Berlin, Germany}
\date{October 2009}
\maketitle

\begin{abstract}
After revisiting some high points of particle physics and QFT of the two
decades from 1960 to 1980, I comment on the work by Jorge Andre Swieca. I
explain how it fits into the quantum field theory during these two decades and
draw attention to its relevance to the ongoing particle physics research. A
particular aim of this article is to direct thr readers mindfulness to the
relevance of what at the time of Swieca was called "the Schwinger Higgs
screening mechanism". which, together with recent ideas which generalize the
concept of gauge theories, has all the ingredients to revolutionize the issue
of gauge theories and the standard model.

\end{abstract}

\section{A brief recollection of quantum field theory in the 60 and 70s}

The years from 1960-1980 mark a high point in particle physics. During these
two decades quantum field theory (QFT) obtained its firm conceptual basis and
its range of applicability to particle physics was considerably expanded to
include all interactions apart from (the still elusive) quantum gravity. This
progress draws mainly on the postwar discovery of perturbative renormalized
quantum electrodynamics (QED) in independent work by Feynman, Schwinger and
Tomonaga, with important conceptual and mathematical additions and refinements
by Dyson. The non-covariant pre-war quantum mechanical perturbation formalism
which can be found in pre 1948 QFT textbooks (Heitler, Wenzel) was ill-suited
for going beyond tree diagrams; it was getting unmanageable for processes
involving interaction-induced vacuum polarization (loop diagrams), of which
some consequences became experimentally accessible shortly after the second
world war. The observational verification of these effects was the entrance of
QFT into the pantheon of established physical theories; in fact the new and
for many (but not all) measurements stupendously precise and successful
covariant formulation of QFT which led to the \textit{Standard Model} placed
it into a very distinguished position within that pantheon.

The progress was foremost methodological. It was not necessary to undergo a
new conceptual revolution to achieve these surprising new results.
Renormalized QED confirmed the conceptual innovations of the true
revolutionary protagonists of QFT (Dirac, Jordan) which were achieved two
decades before. But without the convincing experimental confirmations of the
effects of vacuum polarization in QED, QFT may have disappeared for some time
from the screen of particle physics, and the wildly speculative and metaphoric
attempts trying to exorcise the "ultraviolet catastrophe" may have continued
well into the 50s. \ By preventing such a scenario, the protagonists of
renormalization theory saved QFT and made it fit for still going innovations.

The young avant-garde of the post-war years in particle theory did not set out
to become revolutionaries. Their resounding success, for which three of them
received the Nobel prize, resulted from their innovative and often technically
quite demanding computations which rendered obsolete the prior wild
speculations about the ultraviolet catastrophe of their more "revolutionary"
predecessors (who often preferred speculation over calculation). They
established the correctness of the principles on which the true
revolutionaries of the 20s and 30s founded QFT. \ Without their achievements
in QED the later discovery of the Standard Model would have hardly been
possible and the conceptual confusion, which is characteristic for large parts
of contemporary particle theory, would have arrived much earlier (and even
without the intervention of string theory).

The situation continued up to the end of the 70s. After 1980 theoretical
progress about the Standard Model gradually entered an era of stagnation and
part of the particle physics community, spoiled by almost 4 decades of
continuous success of largely simple-minded ideas invented a new research
subject where one could be "revolutionary" and dream about a theory of
everything (TOE). It is certainly interesting and important to analyze the
reasons for the decline of particle physics in detail, but this is not the
intention of this article. To the contrary, here we want to show on hand of
some typical concrete illustrations how some valuable ideas got lost and how
their resumption could lead out of the present stalemate.

Different from the rather short-lived ultraviolet crisis, the crisis
associated with the dominance of a TOE in form of the superstring already
lasts many decades and there is no end in sight since all the competent
potential critics who enjoy the general esteem of the particle physics
community are either gone or silent\footnote{The silence does not mean
consent; for example Steve Weinberg "voted with with his feet" against the new
turn in particle theory already more than 20 years ago.}.

Apart from the magnitude and the number of involved researchers and
publications, the present situation is vaguely reminiscent of a period of
frame of mind when some physicists, including Heisenberg, tried to cure the
"ultraviolet problem" of QFT prior to renormalized perturbation theory by
invoking speculative ideas without comprehensible connections to QFT. But the
number of physicists working on speculative problems (instead of extending the
conceptual range of QFT or searching for a more appropriate computational
method which is more faithful to the underlying conceptual structure of QFT)
was comparatively smaller at that time; in addition the "ultraviolet
catastrophe era" did not last much longer than a decade, not enough time to
cause any rupture or long-lasting mark.

This time the situation is much more serious. Perhaps the most lasting damage
consists in the fact that an enormous amount of knowledge has been lost. In
fact, as previously mentioned, the main motivation for this essay is to bring
back some knowledge and a frame of discourse in which some of the old cut off
ideas can be adapted to the new situation.

Three decades of string theory since 1980 have left their mark on particle
physics. One can dispute its scientific impact, \ but its influence on the
sociology of science, in particular on particle physics, is beyond question.
The several decades lasting dominance of the idea of finding a TOE has created
a community of specialists who lack the broad knowledge about particle physics
of earlier preelectronic times which makes my task to counteract these
tendencies by recreating some of the lost ideas following the path of
contributions by J. A. Swieca quite challenging.

In any case the sociological and intellectual situation in particle physics
during the two decades 1960-1980 was very different from how it developed
afterwards. The main distinction to the present is that there was more
criticism, including auto-criticism. This was considered an asset because, as
the ultraviolet episode had shown without a strong counterbalance to the
necessary speculative frontiers, particle theory would go astray. Any
speculative jumps into the "blue yonder" were usually done from a conceptually
solid platform so that in case of failure of the incursion, there was always
the return option and the possibility to investigate a slightly different
direction. This option does not exist anymore in string theory; to where could
a string theorist return ? To the dual model, the S-matrix bootstrap or to
that kind of string-influenced QFT of many articles and modern text books in
which QFT features as an effective string theory ? Physicists in those days
had a much greater awareness that a delicate equilibrium between innovative
ideas and a critical mind is the precondition for progress in particle physics.

Sometimes the critical and the innovative abilities came together; A famous
figure who combined both qualities in his persona was Wolfgang Pauli. His
impressive creativity stood next to his cutting criticism, which if necessary,
he did not spare against himself\footnote{After having worked for almost two
years together with Heisenberg on the ill-fated "nonlinear spinor theory" (a
kind of precursor of quarks in which all the observed nuclear particles are
composites of a fundamental spinor field), Pauli abruptly (without looking for
excuses) abandoned and criticized these attempts after Feynman showed him the
fallacies.}. The sociology in particle physics has changed; nowadays it is not
only the predictive power and the theoretical conclusiveness which determines
the status of a theory, but also the market value and its accretion in a
globalized world. The 60s and 70s were the high point of what in Germanic
languages can be expressed in terms of one word, the
"Streitkultur"\footnote{The arena of the Streitkultur of the 60s and 70s were
conferences and was reflected in many conference reports, sometimes even in
regular articles \cite{Jost}.}.

In stating such observations one should be careful of not become accused to
glorify the past at the cost of the present. There was a critical situation in
the two decades before the 80s which resulted from a clash between those who
advocated a pure S-matrix approach and others who considered the S-matrix and
the analytic properties of scattering amplitudes as the litmus test of the
principles on which QFT outside of perturbation theory (hence in particular of
strong interactions) is founded. In addition to the general properties of
quantum theory these were specifically causal locality and the closely related
Poincar\'{e} covariance and energy positivity.

The unfortunate ideologic attitude of the S-matrix purists led to a
confrontation of the S-matrix bootstrap with QFT at the end of the 60s. It was
a struggle about a S-matrix approach cleansed of all field theoretic aspects;
the fervor of its proponents was certainly related to the fact that in those
days for the first time that magic idea of a unique theory of everything (TOE)
entered the discussion (the unique S-matrix bootstrap of all forces apart from
gravity \cite{Chew}). On the other side of the fence there was renormalized
perturbative QFT enriched by the (nonperturbative) LSZ/Haag-Ruelle scattering
theory \cite{Haag} which was shown to be a \textit{structural consequence of
the principles} underlying QFT. The ideological fervor found its strongest
expression in conference reports were the S-matrix bootstrap proponents felt
more free to celebrate what they perceived as their (premature) victory over QFT.

The counter message from quantum field theorists essentially amounted to
remind particle physicists that even if one's main interest are the
on-mass-shell observables as the scattering amplitudes and formfactors, one
still needs the interpolating fields as the carriers of the locality principle
to implement the desired S-matrix- and formfactor- properties by deriving them
from the basic spectral and causality properties of particle physics. Indeed
the bootstrap program lacked even the means to implement its most celebrated
addition to particle theory, the crossing property (which follows from QFT
\cite{crossing}) and, which is a serious flaw, it never addressed those
requirements which \textit{macro-causality} imposes on any multi-particle
S-matrix \cite{interface}. These properties were first listed by Stueckelberg
who also used them to criticize the previously mentioned Heisenberg S-matrix
proposal\footnote{The first attempt to bypass QFT and formulate particle
physics solely in terms of the S-matrix is due to Heisenberg \cite{Hei}. He
wrote wrote down models of unitary Poincar\'{e}-invariant operators. His
proposal fulfilled the spacelike cluster factorization property but violated
Stueckelberg's timelike "causal rescattering" requirement. Both properties are
aspects of "macro-causality" and \ can be formulated and argued (different
from the later "crossing property") in terms of particles without fields.};
they basically consisted in the spacelike cluster factorization and the
absence of timelike precursors (the macro-causal origin of the Feynman
$i\varepsilon$ prescription).

The ferrocity of the struggle on the side of the S-matrix purist against QFT
is hard to understand in retrospect, but the future of particle theory could
have taken another turn if it would not have been for the saving grace of
nonabelian gauge theory which led to a surge in particle theory, starting at
the beginning of the 70s and which sent the first TOE (everything except
gravity) in form of the S-matrix bootstrap into the dustbin of history.

There is however a somewhat ironic epilogue to this second crisis (remember
the first was the "ultraviolet catastrophe"). Those properties as unitarity,
invariance and the crossing property which some years later permitted a
mathematically clear formulation and implementation in the context of two
dimensional factorizing models (section 7) were completely sound; in
connection with the \textit{nuclear democracy} setting of bound-states they
turned out to be extraordinarily successful within the setting of
two-dimensional factorizing models \cite{crossing}. Instead of the expected
TOE from the metaphoric bootstrap idea, one obtained a rich nonperturbative
world of an infinite number of concrete models which, although exhibiting no
on-shell particle creation, share with general interacting models the presence
of infinite vacuum polarization clouds. In other words instead of one theory
of \textit{everything} one obtained an infinite family of models which
constitute a theoretical laboratory for learning \textit{something} about
unknown aspects of QFT. In view of the fact that this is the first
nonperturbative construction of a family of interacting models with a
mathematical existence proof, this is not a small achievement.

But of couse a new constrution of models of QFT based on bootstrap ideas was
not at all what the protagonists of the S-matrix bootstrap had in mind; there
idea was to do away with QFT once and for all. I will return to the issue of
factorizing models in connection with Swieca's contributions in the later part
of the essay (section 7).

The demise of the S-matrix bootstrap was the beginning of a serious crisis,
but as it happens in real life, this was for a \ long time to come not
perceived as such. The difficulty with implementing the crossing property,
which mixes the one-particle contributions with those of the scattering
continuum after analytic continuation and whose true conceptual origin has
only been understood recently \cite{crossing}, led Veneziano \cite{Ven} to the
\textit{duality} requirement in which a formal crossing property (not the QFT
crossing) was obtained with the help of infinitely many intermediate
one-particle states. This dual S-matrix Ansatz led eventually to the string
theory of the 80s and became a fashionable topic of present day particle
theory which achieved its dominating position without observational and
conceptual credentials only by the faith of its reputable protagonists; as a
result of its bizarre consequences it also entered deeply into the popular
science culture \cite{Green}.

After this interlude on developments outside and often in antagonism to QFT,
it is time to look more closely at the aftermath of perturbative
renormalization theory, one of the area which attracted Swieca's interest.

With an enhanced confidence in the physical relevance of QFT, it was possible
at the beginning of the 60s to revisit some old problems of QFT which, despite
the new methodological progress of renormalization theory, did not loose any
of their conceptual challenge. One of those was the problem of "particles
versus fields"\footnote{This particle-field relation is a problem in the
setting of field theoretic localization and the associated vacuum
polarization. It should not be confused with the particle-wave duality of QM,
which is related to the uncertainty relation and Born's probabilistic
definition of localization \cite{interface}.}. Already in the 30's, shortly
after the discovery of vacuum polarization which was first noticed in studying
conserved currents of charged free fields by Heisenberg, Furry and Oppenheimer
\cite{Fu-Op} perceived to their surprise that interacting Lagrangian fields
applied to the vacuum inevitably generate infinite (increasing with
perturbative order) vacuum-polarization "clouds" in addition to the desired
one-particle component.

It maybe helpful to present some details of these observations within a modern
conceptual setting. Heisenberg's observation in modern terminology was that a
"partial charge" in a spatial sphere of radius $\not R $ and volume $V$
\begin{align}
Q_{V}  &  =\int_{V}j_{0}(x,t)d^{3}x\label{Hei}\\
j_{\mu}(x,t)  &  =:\phi^{\ast}(x,t)\overset{\leftrightarrow}{\partial}_{\mu
}\phi(x,t):\nonumber
\end{align}
diverges quadratically and he realized that such an object, which would be
perfectly finite (it even vanishes on the ground state) in QM, must diverge in
QFT as a result of the presence of particle-antiparticle creation operators
whose appearance is characteristic for QFT in comparison to QM. The occurrence
of such particle-antiparticle pairs is what is meant by the terminology
\textit{vacuum polarization}. In the interacting case (the one studied by
Furry and Oppenheimer) different from the free field composite (\ref{Hei}),
the number of such pairs is actually infinite, in which case on speaks of a
vacuum polarization "cloud". In both cases the vacuum polarization
contribution disappears in the limit $V\rightarrow\infty$ so that
$Q\Omega=\lim_{V\rightarrow\infty}Q_{V}\Omega=0$ i.e. the charge of the vacuum
is zero as expected.

Infinities in QFT inevitably indicate that certain concepts have not been
properly understood. In the case at hand it is the singular nature of fields
and currents. Very different from classical fields, covariant fields of QFT
are "operator-valued" distributions i.e. objects which only after smearing
with Schwartz test functions become (mostly unbounded) operators. The
definition of a partial charge with finite vacuum polarization, which has the
property of loosing its vacuum polarization cloud in an appropriately global
limit, was first formulated by Kastler, Robinson and Swieca \cite{KRS}.

We will return to these issues in section 4 where some of the mathematical
details will be presented. There we will explain also how in terms of
spacetime smearing one defines a (dimensionless) partial charge $Q_{R,\Delta
R}$ in a sphere of radius $R$ with a shell of thickness $\Delta R$ for the
vacuum polarization cloud to attenuate in such a way that the norm of the
state $Q_{R,\Delta R}\Omega$ follows (apart from a logarithmic correction) for
$\Delta R\rightarrow0$ a dimensionless area law $area/\left(  \Delta R\right)
^{2}.$

The area behavior of the vacuum polarization of a partial charge preempts the
behavior of the dimensionless \textit{localization entropy} which, as a result
of the shared vacuum polarization aspect, also obeys such an area law. In the
latter case one cannot delegate the problem to the use of distribution theory
since, unlike the partial charge, the localization entropy has no
representation in terms of testfunction smearing.

Hence QFT contains some quantities which, in distinction to QM, approach
infinity in the limit of sharp boundaries. From a conceptual point of view
this is not much different to the volume divergence of in quantum statistical
mechanics, in fact there are rather convincing arguments that both (heat bath
and localization thermality) are related \cite{BMS}.

In most articles a momentum space cutoff is introduced when it comes to these
sharply localized quantities. But this is awkward from a conceptual point of
view because the word "cutoff" is used to express the limitation of the
physical validity of a theory at very high energies which often becomes
confused with the idea that the model does not exist and should be viewed as
the effective low energy approximation of a yet unknown theory.

Whereas it maybe very well true that the theory a mathematical model of
reality theory beyond a certain range may loose its physical validity, one
cannot blame this on the vacuum polarization clouds whose divergence in the
limit of sharp localization is a consequence of the causal locality principle
and as such it is independent of the nature of the interaction.

To exhibit these consequences of sharp localization one often needs
sophisticated mathematical tools. One such mathematical addition which was not
yet available at the time of Swieca is the "split property" \cite{Haag}. This
allows to relate the standard heat bath thermal behavior with that caused by
localization. It suggests that the volume divergence of the former and a
(logarithmically modified) dimensionless (quadratically divergent with the
inverse thickness of the polarization cloud) area law have a common origin
\cite{BMS}. These are objective divergeces in QFT and any manipulations with
cut-off would destroy the setting of QFT in an uncontrollable way. Swieca did
not have such powerful new concepts, but the mathematical control of vacuum
polarization in establishing the connection between conserved currents and
would-be quantum Noether charges in his work set the standards for dealing
with localization-caused vacuum polarization at that time.

Here some additional historical remarks about the use of \textit{distribution
theory} in QFT are in order. Already in the early 50s it became clear that
without its use QFT would remain in an unclear metaphoric state, consisting
mainly of computational recipes involving Dirac's delta functions and its
derivatives and a confused conceptual situation without clear separation
between genuine ultraviolet divergences and problems caused by a lack of
appreciation the intrinsic singular nature of field "operators". Even nowadays
there are pockets of resistance i.e. physicists who continue to talk about
ultraviolet divergences and how to get rid of them by renormalization theory
instead of how to formulate the latter in the setting of operator-valued
distributions together with the principles of causal locality as first
formulated by Wightman \cite{Haag}. But besides the mentioned divergences
caused in partial charges and localization entropy in the limit of sharp
localization, there are no intrinsic divergences in QFT; distribution theory
may not be sufficient to see this, but it is certainly necessary.

Distribution theory became after group theory the main mathematical tool for
particle theory and its use by quantum field theorists begun in the 50s.
Swieca's adviser at the University of Sao Paulo, Werner Guettinger, was a
particle physicist in the forefront of this trend. As a result of a close
cooperation with mathematicians (including Grothendiek, Dieudonn\'{e} and
Schwartz \footnote{When I came to the USP for the first time in 1968, there
were courses on distribution theory in the physics department given by a
former PhD student of Laurant Schwartz.}) from France; particle theory at the
University of Sao Paulo profited quite early from this offer. With Jos\'{e}
Giambiagi, these new mathematical tools became also known in Argentina at a
quite early time.

The ubiquitous presence of polarization clouds in problems of quantum field
theoretic localization required a drastic conceptual revision of what one has
learned about the relation between particles and fields in QM where (using the
second quantization setting) the application of the elementary basic field to
the vacuum generates a one-particle state, whereas the application of
appropriately (with the help of bound state wave functions) smeared products
of the basic field leads to a bound state. In interacting QFTs the presence of
infinite vacuum polarization clouds make it impossible to create a
one-particle state without an attached polarization cloud by applying a field
(more generally a compactly localized operator) to the vacuum.

Although the role of vacuum polarization in separating the structure of QFT
from that of QM had its historical roots in the relation of global charges
with conserved currents, the vacuum polarization aspects of QFT pervades
almost every issue. This can be nicely illustrated in terms of formfactors. A
\textit{formfactor} is a general terminology used for matrix elements of a
field between "bra" states, consisting of say $n-k\geq0$ outgoing particles,
and $k$ incoming particles in "ket" states\footnote{Here we make the standard
assumption of scattering theory, namely the validity of \textit{asymptotic
copleteness}. In the absence of zero mass it is not only valid in
paerturbation theory, but it also has been verified in exactly solved
factorizing models (section 7)..}. Taking the simplest case of a scalar field
$A(x)$ between spinless states of one species it reads%
\begin{align}
&  ^{out}\left\langle p_{k+1},...p_{n}\left\vert A(0)\right\vert
p_{1}...,p_{k-1},p_{k}\right\rangle ^{in}\label{cross}\\
&  =~^{out}\left\langle -p_{k}^{c},p_{k+1},...p_{n}\left\vert A(0)\right\vert
p_{1}...,p_{k-1}\right\rangle ^{in}+p_{k}^{c}-contr\nonumber
\end{align}
in words the incoming momentum $p_{k}$ is "crossed" into the outgoing
$-p_{k}^{c},$ where the c over the momentum indicates that the particle has
been crossed into its antiparticle and the unphysical (negative mass-shell)
momentum turns out to be defined in terms of analytic continuation properties
(which in turn follow from the modular localization properties of QFT
\cite{crossing}). The $p_{k}^{c}-contr$ are contraction terms i.e. delta
functions from inner products $\left\langle p_{k}|p_{l}\right\rangle ,$
$k+1\leq l\leq n$ multiplied with lower formfactors which are there in order
to preserve analyticity (they compensate a nonanalytic delta function
contribution to the first term).

The relation (\ref{cross}) would be physically void if it would not come with
an assertion of analyticity which connects the unphysical backward mass shell
momentum with its physical counterpart. This kind of crossing property permits
to reduce all formfactors of a localized operator $A$ to the particle
components of a "bang" on the vacuum $A\Omega,$ the name for a sharp acoustic
excitation here serves as a metaphor for local excitation of the vacuum which
contains the full energy-momentum spectrum up to infinity. In this setting the
various components of the vacuum polarization cloud associated with the
localized operator $A$ are described by
\begin{equation}
\left\langle p_{1},...p_{n}\left\vert A\right\vert \Omega\right\rangle
,~n=1,2,....
\end{equation}
I prefer this terminology of a bang on the vacuum to that of a "broiling soup"
which for short times is allowed to violate the energy-momentum conservation
law. Admittedly both pictures use a somewhat metaphoric terminology, but the
former has at least a precise physical content. It also gives a concrete
meaning to the adaptation of \textit{Murphy's law} to particle physics (for a
precise mathematical formulation see \cite{crossing}):

\textit{Claim}: \textit{Localized states in an interacting QFT which are not
forbidden (by superselection rules) to mutually couple, do indeed
couple\footnote{For the present purpose absense of coupling means simply
orthogonality.}.}

Whether one considers this a "benevolent" or the better known troublesome form
of Murphy's law depends on ones aim; if one wants to apply operator methods
from QM\footnote{This will inevitably lead to infinities and cutoffs which
have not only no intrinsic meaning, but also convert the originally local
theory into something which apart from mathematical problems has no known
conceptual position.} (i.e. outside the range of vacuum polarization and
Murphy's law) to QFT, one is in for serious trouble; if on the other hand one
looks for a framework of a fundamental theory in which the different models
are realizations of a few underlying physical principles, Murphy's law is an
unmerited blessing. There is no other theory in quantum physics in which the
observational wealth can be reduced to the realization of one principle of
causal localization and (at least at zero temperature) the closely related
covariance and positive energy requirement.

After this short step into the presence, the historical outline about the
situation which Swieca encountered in the 60s will be continued.

The important step in the post renormalization period of QFT was the
clarification of the field-particle dichotomy. The ubiquitous presence of
vacuum polarization clouds prohibits any naive coexistence of particles and
fields as one is used to from QM; in spite of the central role of the notion
of a particle in measurements, the ontological status of particles in QFT is
considerably weakened as compared to QM. The derivation of the S-matrix from
the large-time asymptotic behavior of fields \cite{Haag} was a great leap
forward, since at least the large time asymptotic region was protected against
vacuum polarization clouds. It became clear that, quite different what one
expects on the basis of an analogy with QM, multiparticle states only acquire
a frame-independent intrinsic meaning through scattering theory i.e. at
asymptotic large times. Sharply spacetime-localized states in interacting
theories always contain infinite vacuum polarization clouds and their presence
is the most characteristic property of QFT\footnote{For the (later mentioned)
d=1+1 factorizing model the S-matrix is purely elastic but despite the absence
of on-shell particle creation the interaction-caused vacuum polarization
clouds ("virtual" or off-shell particle creation) are fully present.}; their
mathematical control require conceptually quite challenging ideas.

This research also led to a better understanding of the relation of the vacuum
polarization clouds as intrinsic \textit{local} indicators of the presence or
absence of interactions\footnote{The earliest such theorem (the Jost-Schroer
theorem \cite{St-Wi}) states that the absence of a vacuum polarization cloud
in a "one-field state" characterizes a theory generated by free fields. in . A
recent more powerful generalization characterises the absence of interactions
in cases of much weaker localization properties \cite{Jens}.}. Last not least,
the S-matrix aspects of QFT also led to a re-appraisal of Wigner's 1939
intrinsic representation theoretical classification of positive energy
irreducible representations of the Poincar\'{e} group as an intrinsic (and
unique) way of characterizing particles which is conceptually superior to the
description in terms of linear hyperbolic covariant (spinorial) field
equations. Whereas the latter is highly non-unique (for a given physical spin
there is always an infinity of spinorial wave functions), Wigner's setting is
unique. Scattering theory is based on the idea that every state under
large-time asymptotic interpretation is a superposition of n-fold tensor
products of Wigner representations. Without the asymptotic stability
properties of n-fold particle localization, it is not possible to formulate
scattering theory of particles within the setting of QFT\footnote{These ideas
about the particle-field relation appear for the first time in \cite{Ha-Sw}.}.

The old pre-renormalization struggle with ultraviolet divergences came to an
end when the message that pointlike quantum fields are by their very nature
rather singular objects which required testfunction smearing was headed also
in n-th order perturbation theory where, together with causal locality, it led
to the statement that the time-ordered correlation functions can be determined
recursively from their lower order contributions up to a new delta function
term (which only contributes on the total diagonal i.e. if all localization
points coalesce) \cite{E-G}. Together with a theorem about the structure of
pointlike composites of the free fields \cite{St-Wi} this fixes the form of
the "counter-terms" of renormalization theory. If the iteratively determined
part of the correlation function (together with an restriction on the
singularity degree of counterterms) has a short distance behavior which can be
majorized by a certain short distance scaling degree independent of the order
of perturbation theory (determined by the power-counting limit), the resulting
perturbation theory depends only on finitely many parameters and is referred
to as "renormalizable".

It is believed that only renormalizable theories have a conceptional
mathematical reality outside perturbation theory\footnote{The perturbative
series in QFT are all known to diverge; so RPT has no conceptional
significance for the existence of a model (a unique situation which has no
counterpart in other branches of theoretical physics). However in the special
setting of two-dimensional factorizing models (section 6) all the exactly
solved models are also renormalizable in the perturbative sense.}. This
completely finite and cut-off independent formulation exists thanks to Epstein
and Glaser \cite{E-G} since 1973 and became the preferred renormalization
setting for those who consider renormalization a foundational problem which
should not be left to a set of computational recipes leading to ultraviolet
divergencies and cutoffs. The Epstein-Glaser perturbation theory is not the
only cutoff-free formulation, but it is the one with the clearest relation to
the underlying locality and positive energy spectrum condition and on the
other hand with the greatest distance to the quantization parallelism of
classical field theory in form of the Lagrangian approach.

In these remarks I tried to recapture the Zeitgeist and the scene which Jorge
Andr\'{e} Swieca encountered in the beginning of the 60s when he entered
particle physics and which he, together with others, shaped during the two
decades of his scientific activity.

\section{The Haag-Swieca work on phasespace degrees of freedom}

What makes Jorge Andre Swieca an interesting figure in connection with a
review of particle physics of the 60s and 70s is that, although he started his
career in the highly conceptual-mathematical oriented group of researchers on
local quantum physics (LQP) which formed at the beginning of the 60s at the
University of Illinois in Champaign-Urbana around Rudolf Haag, he belonged to
the very few individuals from that kind of background who used their basic
knowledge not only to advance the conceptual framework of LQP, but also in
order to solve problems closer to the ongoing particle research. This allowed
him to have a more critical and in many cases also more profound access than
others to solutions. It is the purpose of this essay to exemplify this by
reminding the reader of some of those ideas.

To follow the ideas of Swieca is also instructive from another point of view.
The subject of spontaneously broken symmetries and the Schwinger-Higgs
screening mechanism were certainly quite competitive subjects at the time, but
Swieca's contributions to these topics were completely original and somewhat
different. Revisiting these subjects with a modern hindsight will be a new
experience for many in the younger generation.

This is in particular true about his first paper, after having obtained his
Ph.D. at the University of Sao Paulo in 1964, a paper written together with
Rudolf Haag at the University of Illinois in 1963 under the very ambitious
title "when does a Quantum Field Theory describe Particles?" \cite{Ha-Sw}. The
authors aim at a completely intrinsic conceptual understanding of particles in
terms of causal localization properties of fields and the closely related
positive energy condition, a very reasonable strategy in a theory whose main
distinction within the general setting of quantum physics is the relativistic
localization\footnote{This was not the impression one was getting from most
textbooks. For this reason the terminology \textit{local quantum physics}
(LQP) was used whenever the underlying principles and their consequences and
not the (perturbative) quantization were the main focus of interest.}. The
quantization of theories with a maximal velocity (QFT) as compared with those
without (QM) is much more restrictive; whereas QM either nonrelativistic or
its relativistic DPI version \cite{interface}) is not subject to restrictions
(beyond the requirement that the interaction potentials are not too long
ranged), interactions in QFT are subject to more restrictive conditions
resulting from locality. According to renormalizable Lagrangian quantization
the local couplings only consist of finitely many coupling parameters and
accepting the widespread belief that higher spin interactions are ill-defined
(nonrenormalizable), there exists only a finite number of renormalizable
interacting models.

The remarkable finding in the Haag-Swieca paper was that the locality
principle in QFT requires more phase space degrees of freedom than the
well-known quantum mechanical law of a finite number of degrees of freedom per
unit phase space cell in QM; in fact the H-S result was that although the
phase space degrees of freedom was infinite, its cardinality does not surpass
that of a compact set. It was Haag's dream, ever since the birth of local
quantum physics (LQP, often referred to as algebraic QFT) at the end of the
50s, that relativistic particles, as first intrinsically (i.e. without use of
classical quantization-parallelism) classified by Eugene Wigner, are the
asymptotic stable carriers of the locality principle. Whereas in a given
theory there are myriads of fields ("interpolating" fields in the terminology
of LSZ) which, apart from being carriers of conserved charges have a fleeting
observational content, particles are stable autonomous, unique elementary
objects in the Hilbert space whose only slightly metaphoric aspect is that
they only show in asymptotic events. Fields are the carriers of the locality
principle in finite spacetime regions whereas the particle states only
manifest themselves observational at asymptotically large times.

Assuming the existence of an spectrally isolated one-particle state (the
mass-gap assumption), Haag's idea that the spectral and locality properties
are sufficient to derive the LSZ asymptotic condition was beautifully
vindicated. But the title of the Haag-Swieca paper was pointing into the
direction of something more ambitious project since the new aim was to derive
the very existence of one particle states. Although there is no definite
answer up to this day to the central question which these authors ask in the
title of their paper, the richness of the research it led to is quite impressive.

According to my best knowledge this paper is the first in which the difference
between the quantum mechanical and the quantum field theoretical concept of
\textit{phase space} in QFT is seriously addressed. Whereas, as mentioned, in
QM the number of quantum states which can occupy a finite phase space region
$\Omega$ is finite, namely maximally $\Omega/\hslash^{3},$ it was known that
(in the case of free fields) the number of states below a certain energy and
localized in a compact spacetime region $\mathcal{O}$ is still infinite, even
if one, following Haag and Swieca, circumvents the prerequisites of the
Reeh-Schlieder theorem\footnote{The Reeh-Schlieder theorem \cite{Haag} states
that the family of state vectors, obtained by applying smeared fields with
test functions sopported in a given space time region, is dense in the Hilbert
space. This initiated many discussions since it defies quantum mechanical
intuition.} by admitting only such state-creating operators $Q$ which are from
a subset of the local algebra $\mathcal{A(O)}$ consisting of all $O$-localized
operators whose norm is below a certain bound on the vacuum$~\Omega~$namely
(copying from their paper)%
\begin{equation}
\left\Vert Q\right\Vert \leq e^{\kappa r}\left\Vert Q\Omega\right\Vert
\end{equation}
with $\kappa~$= smallest mass and $r$ the radius of a (without loss of
generality) spacetime double cone $\mathcal{O}$.

Calculations for free fields led Haag and Swieca to the result that, although
the number of states in a finite phase space region (finite spacetime
localization and finite energy) is really infinite, it is "essentially finite"
in the sense of being compact i.e. a set whose cardinality of phasespace
states deviates only mildly from the quantum mechanical finiteness per phase
space cell. There were reasons to believe that interactions did not not change
the situation and therefore the authors expected that their compactness
criterion may be a good starting point for understanding the local origin of
the one-particle structure and the asymptotic large time stability of n-fold
localized particle states. Their most ambitious aim was to find an answer to
the crucial question what properties of local fields lead to
\textit{asymptotic completeness} which is the assertion that every state in
the theory can be represented as a superpositions of multi-particle states;
this is a problem which was left open by the LSZ-Haag-Ruelle scattering
theory. Haag and Swieca did not quite achieve this (see also \cite{Enss}), and
the derivation of particle properties from local aspects of fields has
remained an ambition of fundamental research up to this day.

This is not surprising because in contrast to QM a multiparticle state at
finite times becomes a meaningless concept in the presence of
interactions\footnote{Even the existence of a compactly localized one-particle
state with no additional vacuum polarization admixture is inconsistent with
the presence of interactions. Only for the noncompact wedge regions this is
possible; but even in this case the domains of definition of such vacuum
\textbf{p}olarization-\textbf{f}ree-\textbf{g}enerators (PFGs) have very
restrictive properties.}; from the times of Furry and Oppenheimer it was
already known that it is \textit{not possible to locally create a pure
one-particle state} without the admixture of interaction-induced vacuum
polarization clouds (formed from particles-antiparticle pairs). In other
words, although states with a prescribed number of particles exist in the
Hilbert space, no such state can be locally generated, In the presence of
interactions there exists a sharp antagonism of the notion of particles with
the localization inherent in QFT. For this reason Haag and Swieca take great
care for defining n-particle states in terms of asymptotic counter-coincidence
arrangements which they relate to the representation theoretic (Wigner's
Poincar\'{e} representation theory) tensor product structure which is the only
consistent and unique way of avoids contradictions of massive
particles\footnote{In the presence of zero mass one may end up with
infraparticles which require a different scattering framework.} with field
localization in the presence of interactions. Only in free field theories
there is a close relation between particles and smeared free fields, with the
mass-shell projection of the spacetime smearing function being the particle
wave function.

From a contemporary point of view the reason behind this contrast is the
substantial conceptual difference between the quantum mechanical "\textit{Born
localization}" (in the relativistic context the Born-Newton-Wigner (BNW)
localization) which formed our physical and mathematical notion of particles
as opposed to the field theoretic \textit{modular localization; }for a recent
treatment of this subject \cite{MSY}\cite{interface}). The way modular
localization increases the state density in the phasespace of QFT as compared
to that in QM is through the persistent presence of vacuum polarizations at
the horizon (the causal boundary) of a localization region. Relativity in the
form of the covariant representation theory alone is not sufficient for the
occurrence of vacuum polarization, as the existence of "direct particle
interaction" shows \cite{interface}. Their cardinality of phasespace degrees
of freedom is quantum mechanical i.e. a finite number per phasespace cell. On
the other hand every covariant quantum theory with a sharply defined maximal
velocity will lead to localization-caused polarization clouds and define a
QFT, even if it cannot be viewed as coming from a Lagrangian.

Examples are certain generalized free fields which, as the result of their
plethora of degrees of freedom, are pathological since they cause violation of
the \textit{timeslice property} (the quantum counterpart of causal
propagation) and do not pass the Haag-Swieca phase-space test either
\cite{Ha-Sw}. \ 

Later other authors re-investigated this problem and succeeded to sharpen the
estimates by showing that via the use of a slightly different formulation one
could replace compactness by \textit{nuclearity.} Compact subsets in infinite
dimensional Hilbert spaces are smaller than bounded sets and nuclear sets are
even slightly more meagre.

This important step was taken two decades after the Haag-Swieca paper by
Buchholz and Wichmann \cite{Bu-Wi}. This more stringent (but harder to
establish) phase space property of QFT went a long way to clarify some thermal
aspects of QFT. Roughly speaking it assured the existence of a thermal
equilibrium KMS state once one knows the local observables in their vacuum
representation \cite{Bu-Ju}. Since the thermal representation is unitarily
inequivalent to the vacuum representation, this is not as simple as its sounds.

It is interesting to take a more detailed look of what was accomplished. The
map whose nuclearity is under discussion is a map from operators in an
operator algebra of local observables $\mathcal{A(O})$ to states in the
Hilbert space $H.$ More precisely their sharpened version states that the set
of state vectors obtained by applying the energy damping operator $e^{-\beta
H}$ to the local algebra $\mathcal{A(O})$ defines a nuclear map $\Theta$%
\begin{equation}
\Theta_{\mathcal{O},\beta}:\mathcal{A(O})\rightarrow H,\ \ \ \ A\rightarrow
exp(-\beta H)A\Omega,~A\in\mathcal{A(O})
\end{equation}
A set of states is called \textit{nuclear} if it can be included in the range
of a trace-class operator. A nuclear set in a Hilbert space $H$ is a set which
is dominated by the range of a trace-class operator. Since a trace class
operator is always compact, nuclear sets are a fortiori compact.

A more intrinsic implementation of the phase space idea, which uses only
objects which refer to local algebras, consists in employing instead of the
exponential damping factor involving the Hamiltonian of the modular operator
$\Delta_{\overset{\symbol{126}}{\mathcal{O}}}$ associated with a slightly
bigger spacetime region $\overset{\symbol{126}}{\mathcal{O}}~\supset
~\mathcal{O}$ \cite{Haag}. The modular operator is a mathematical object which
is directly related to the algebra $\mathcal{A(O)}.$

For "pathological" field models, as the generalized free field considered by
Haag and Swieca (in order to show that a reasonable phase space behavior is
not a consequence of locality and energy-momentum positivity alone), \ thermal
states may either not exist at all or they may lead to a maximal (Hagedorn)
temperature. This is a serious problem in theories with infinite particle
towers as string theory.

Needless to add that the issue is still very much alive, and the original aim
of understanding the role of phase space degrees of freedom in relating
particles and their properties with fields is still on the research agenda, as
a glance at a most recent paper shows \cite{Dyb}. Looking at the introduction
of that paper the author leaves no doubt about where this line of research originated.

The Haag Swieca work belongs to those few papers of the middle of last century
with carry an important legacy since the ideas around the size of the phase
space in QFT, and the subtle consequences for particle physics are still far
from closure.

Although the validity of the asymptotic completeness of particle states has
meanwhile been established for the class of factorizing models \cite{Lech},
the Haag-Swieca quest for a general structural derivation of these properties
from local properties has not yet been accomplished; another indication that
QFT is still a far cry from its closure.

\section{Lost knowledge and the Maldacena conjecture}

The knowledge about the phase space restrictions which distinguish pure
\ mathematical models of QFT (axiomatic QFT) from those with physical
relevance remained limited to the rather small community of LQP. With
increasing frequency since the 80s most particle theorists view QFT basically
as a collection of computational recipes. This would be adequate as one uses
computational tools as scattering theory, Lagrangian quantization, functional
methodes and renormalized perturbation within the boundaries of their
limitations\footnote{One of the limitation is that renormalized perturbation
leads to a divergent series (in fact not even Borel summable) and hence
contains no information about the mathematical existence of a theory.
Nevertheless properties which can be shown in every order of perturbation
theory (vanishing of anomaly coefficients and beta-functions) are believed to
be a structural property of the would-be solution.}. But even though one
cannot decide whether a model associated to Lagrangian quantization exists,
one has all reasons to be quite confident that if it exists it will be a
theory whose degrees of freedom cardinality is that as postulated by Haag and
Swieca. A violation would lead to the nonexistence of temperature states (or
at least to the appearance of a limiting Hagedorn temperature) and the
breakdown of the quantm adaptation of the causal propagation property, thus
leading to a clash with properties attributed to the Lagrangian quantization.
Of course the violation of any of those physical properties does not create
\textit{mathematical} problems.

The loss of knowledge about these subjects of the 60s did not remain without
consequences with respect to the issue of the anti de Sitter-conformal field
theory (AdS$_{n+1}$-CFT$_{n})$ correspondence. It has been known for a long
time that QFTs on n+1 dimensional anti de Sitter spacetime (de Sitter
spacetime with negative curvature) and QFTs on n-dimensional conformal models
on n-dimensional Dirac-Weyl compactified Minkowski space share the same
vacuum-preserving spacetime symmetry group $\widetilde{O(4,2)}$ and, as a
result of the close connection between the concept of localization and
covariance, it appeared plausible that there could be in addition to the
shared spacetime groups also a local correspondence between these two models
in different spacetime dimensions; though it could not extend to the
point-like generating fields simply because there is no invertible pointlike
transformation of a spacetime to a lower dimensional one.

The issue lay dormant for many years, the model only served as a remainder
that the Einstein-Hilbert equations admits solutions with closed timelike
worldlines and hence had to be supplemented by additional requirements which
exclude such "time-machine" solution.

The issue returned when Maldacena \cite{Mal} revived the old idea that the
anti de Sitter spacetime $AdS_{5}$ and conformal quantum field theory in one
less dimension $CFT_{4}$ could share more than just the spacetime symmetry
group $\widetilde{O(4,2)}.$ He put forward the idea that the mathematical
AdS-CFT relation could perhaps lend support to the speculative idea that gauge
theories may be related to some form of spin two gravity theory.

In the context of a supersymmetric N=4 Yang-Mills theory, which was the only
4-dimensional theory for which the vanishing of the Beta function in low
orders had been established, this nourished hopes that the theory may be
conformal invariant and therefore could serve as a candidate in a CFT-AdS
correspondence. On the AdS side the Maldacena conjecture expected a
supersymmetric interaction involving a s=2 symmetric tensor representing a
5-dim. gravitational field.

Two remarks on this conjecture are in order. In the 70s there have been
rigorous and elegant methods \cite{Lo-Sc}\cite{Gomes} to prove the absence of
radiative corrections to certain anomalies as well as of the Beta
function\footnote{The Beta function is known to appear in the trace ot the
enrgy momentum tensor and its vanishing is the prerequisite for conformal
invariance in the sence that the zero mass limit exist and is conformal
invariant.}. They consisted in combining the parametric Callen-Symanzik
equations with the Ward identities in order to abstract an equation for which
the nonvanishing of a certain coefficient in lowest perturbative order already
establishes the identical vanishing of the desired expression to all orders.
Apparently either the knowledge about these techniques have been lost, so that
it becomes a matter of faith to accept the claimed properties from the lowest
order computation.

The second remark which adds weight to the title of this section and which
leads us back to the Haag-Swieca work is the following. Even if one does not
worry about details about the conformal status of the supersymmetric N=4 Yang
Mills theory and the precise nature of the object on the AdS side which
correspond to it, there is the problem to understand why both of the theories
should be physical in the sense of having the physically required phase space
degrees of freedom in their respective spacetime dimensions in the sense of
the previous section. Already simple minded arguments would suggest that
starting from a 5-dimensional \textit{physical} AdS theory and reordering its
degrees of freedom according to the spacetime structure of a 4-dimensional
conformal QFT would lead to an abundance of degrees of freedom i.e. to
precisely such a situation (breakdown of causal propagation) which Haag and
Swieca ruled out by their degrees of freedom criterion. If one starts on the
other hand from a physical $CFT_{4}$ model, the degrees of freedom are too
"anemic" in order to cover the $AdS_{5}$, they will hover near the boundary of
AdS and are unable to "fill" the higher dimensional spacetime in order to
produce a physical QFT. The opposite problem of an "overpopulation" could
threaten the conformal side, not in a mathematical sense but in the sense of
obtaining a physically sick theory.

This simple minded argument has been mathematically established \cite{Reh} and
can be nicely illustrated with a free AdS field where one can explicitly see
that the CFT is a \textit{generalized free field} whose K\"{a}ll\'{e}n-Lehmann
spectral function increases in such a way that it develops those physical
pathologies\footnote{In physical theories the operator algebra of a spacetime
region $\mathcal{O}$ is equal to that of its causal completion $\mathcal{O}%
$\textquotedblright. In the case of presence of too many degrees of freedom
there are "\textit{poltergeist" degrees of freedom entering "sideways": so
that the causally completed algebra becomes bigger }$\mathcal{A(O}%
^{\prime\prime})\supset\mathcal{A(O}).$and the timeslice principle \cite{Haag}
is violated.} which one wanted to avoid with the Haag-Swieca requirement. The
other direction i.e. starting from a conformal free field is a bit tricky
since pointlike AdS fields are in this case not available. Nevertheless there
exists a one-to-one relation between operator algebras localized in certain
spacetime regions and since arbitrary causally closed regions can be obtained
via intersections, this is sufficient to reconconstruct the local net which is
the algebraic replacement for pointlike fields. Following this path, the local
AdS-CFT correspondence as a statement of a structural property, including the
mismatch of degrees of freedom, has been proven by Rehren \cite{Reh}.

Obviously correspondences in different dimensions cannot be formulated between
pointlike fields. They can however be established between algebras associated
to certain noncompact regions and by taking intersections between these
algebras one works one's way down to sharper localized compact localized algebras.

The equality of degrees of freedom in a relation between QFT of different
dimension is related to the equality of the spacetime symmetry groups. It does
not occur in the holographic relation between a QFT in a bulk region and that
of its horizon. Such holographic relations are necessarily \textit{degrees of
freedom reducing holographic projections} accompanied by the reduction of the
symmetry group: of the 10-parametric Poincar\'{e} group in 4 dimensions only a
7-parametric subgroup survives the projection \cite{BMS}. In this case the
cardinality of degrees of freedom on the lightfront corresponds precisely to
what is "physical" in the sense of Haag and Swieca. The holographic projection
onto a lower-dimensional timelike "brane" is analogous to the AdS-CFT case
although the action of the larger spacetime symmetry on the degrees of freedom
of the brane is more tricky since it looses its geometric significance.

The fact that there have been more than 6000 publication on such a relative
narrow subject as the conjecture about AdS$_{5}$-CFT$_{4}$ correspondence on
which, according to the above remarks one anyhow cannot expect a physically
acceptable solution on both sides of the correspondence, is a measure of the
depth of the crisis which particle theory entered when it tried to become the
end of the millennium "theory of everything".

An hypothetical observer returning from the past after a 30 year would perhaps
suspect that the physical facts have undergone a radical change. But the only
aspect which changed is that some people have worked very hard to make
metaphoric arguments more acceptable. Nobody will negate the value of
metaphors in keeping important ideas alive before they have been
mathematically and conceptually secured. However accepting theories with extra
dimensions which are then ordered to curl up and become internal symmetries is
a different cup of tea.

The contrast between the spirit in particle physics at the time of the
Haag-Swieca work and the present Zeitgeist can be condensed into the following
statement. Whereas in earlier times physical arguments served to select
between mathematically consistent possibilities, the present trend is the
converse namely to claim that everything which is mathematically possible
admits a physical realization. Admitting a higher than physical cardinality of
phasespace degrees of freedom as obtained on the CFT side from a physical AdS
model leads to a violation of causal propagation. In more detail, the degrees
of freedom in a spacetime region $\mathcal{O}$ is smaller than in its causal
completion $\mathcal{O}^{\prime\prime}$ i.e. the additional degrees of freedom
have entered sideways in the manner of a "poltergeist". The most outspoken
representative of the viewpoint that every consistent mathematical structure
has a physical realization is Tegmark \cite{Teg}. It is clear that it is more
difficult and time-consuming to do research under the conceptual weight of
physical principles than performing free-wheeling calculations which only have
to be acceptable to community members with a similar level of knowledge and
philosophy about what is physical.

As a result large parts of knowledge have been lost, and with the unshakable
confidence which only an ideology as a TOE supports (but which is alien to the
auto-critical spirit of traditional science), there is in the eyes of many (in
particular among those who entered particle theory after 1980) no virtue to
loose time in studying old ideas while there is the historical chance of
participating in the project of a theory of everything. This explains why
there has been (and still is) this incredible high number of papers on a
subject of only modest physical interest.

The fact that an increasing number of physicists who received their scientific
formation in the shadow of these problems are now referees in formerly
reputable journals is aggravating the situation and does not leave much hope
for the near future. Another detrimental aspect of this situation is that of
prematurely (before a problem has been solved) given awards which confer to
the winners the aura of protection and invulnerability (respected in
particular by referees and editorial boards of journals). This has essentially
destroyed the old "Streitkultur" at the time of the 60/70s which played an
important role to keep particle theory on a high and sound level.

\section{Spontaneously broken symmetries}

A second set of problems which received a lot of attention during the two
decades under discussion was \textit{symmetry and symmetry-breaking}. Both
issues were initially investigated in the Lagrangian quantization setting; the
first presentation of Lagrangian spontaneous symmetry breaking is due to J.
Goldstone \cite{Go}. An older condensed matter version of spontaneous symmetry
breaking in the setting of spin-lattices goes back to Heisenberg and his
theory of ferromagnetism; although as a result of its special nature in solid
state physics it was not perceived as a special illustration of a vastly
general phenomenon in systems with infinite degree of freedoms which includes QFT.

In the Lagrangian setting Goldstone's derivation established the existence of
a symmetry breaking in a particular model and it was left to the reader to
decide take this either as a property of a special model or class of models or
to muster enough faith to belief in a general structural theorem of QFT behind
this observation. In order to prepare the ground for a more autonomous
discussion\footnote{An understanding which does not refere to the way a model
has been constructed but only uses intrinsic properties of its presentation in
terms of expectation values.} it was necessary in a first step to state the
meaning of quantum symmetry in a way that a spontaneous breakdown of such a
symmetry looks like a meaningful natural generalization.

The starting point Swieca and collaborators took was to aim for a precise
definition of the global charge associated with a conserved current that of a
conserved quantum current and its expected role as a generator of a symmetry,
but now with a more precise definition as a spacetime limit of a sequence of
\textit{partial charges} in terms of test function smearing of the zero
component of the current.

The test function smearing of the current tames the vacuum polarization cloud
and for the partial charge contained in the region $\left\vert \mathbf{x}%
\right\vert \leq R$ one defines%
\begin{align}
Q_{R}  &  =j_{0}(f_{R},f_{d})\\
f_{d}  &  =0~~for~\left\vert x_{0}\right\vert \geq d,~f_{R}(\mathbf{x})=%
\begin{array}
[c]{c}%
1~for~\left\vert \mathbf{x}\right\vert <R\\
0~for~\left\vert \mathbf{x}\right\vert \geq R+\varepsilon
\end{array}
\nonumber
\end{align}
Using the conservation law for the current one can then show that on any local
operator $A\in\mathcal{A(O}_{R})$ the commutation relation with $Q_{R}$ is the
same as with the global charge i.e. algebraically there is no divergence
problem of the partial against the global charge. But of course one wants
convergence properties on states and this is where the control of vacuum
expectations is essential; it is also where the difference between bona fide
symmetries and spontaneously broken symmetries are beginning to show up.

The strength of the divergence in the limit $\varepsilon\rightarrow0$ (the
limit of sharp partial charge) can easily be computed from the two point
function of the current and follows, apart from a possible logarithmic
$\varepsilon$-divergence the leading term follows the dimensional
rule\footnote{The only divergence in d=1+1 is logarithmic.}, which for a
dimensionless charge would be $\frac{area}{\varepsilon^{2}}$ with the area
being proportional to $R^{2};$ for the leading term the details of the test
function do not play any role .

The large distance limit in which one expects the global charge to emerge is
very different, in particular the $\varepsilon\rightarrow0$ behavior does not
enter. First one observes that even in the best of all cases (namely the
current associated to a free field) the convergence to the global charge for
$R\rightarrow\infty$ is only in the sense of weak convergence on a dense set
of states. This is easily proven for theories with a mass gap. At this point
there is precisely one step which could spoil the convergence, namely the
presence of a massless and spinless particle which couples to the current and
prevents the weak convergence on the vacuum. This is the famous Goldstone
boson. The proof that the spontaneous breaking requires the presence of a
$\delta(\kappa^{2})$ contribution in the K\"{a}llen-Lehmann function of the
current. A very beautiful proof of this theorem with the help of the
Jost-Lehmann-Dyson representation was given by Ezawa and Swieca \cite{ES}.

Note that the famous field vacuum expectation value (the Nambu-Goldstone
"condensate") is not an intrinsic aspect of spontaneous symmetry breaking but
a technicality of its implementation in certain Lagrangian models. For this
reason one will not find such concepts in structural investigations; as useful
as they may be in model calculation, at the end of the day they disappear from
the observables which are dominated by masses and spin of particles as well as
scattering amplitudes and formfactors of currents. The only intrinsic mark by
which it differs from simply having no symmetry is the appearance of a zero
mass Goldstone Boson which couples in a specific way to the conserved current.

The state of art on symmetry versus spontaneous symmetry breaking of the 60s
can be found in the 1970 Cargese lecture notes by Swieca. In these notes these
ideas are also adapted to nonrelativistic many body problems. In that case the
vacuum polarization effects are absent and the locality principle is replaced
by assumptions about the range of interactions.

In the years after 1970 there were several refinements.

Since the problem of conserved currents was the first in the history of QFT
which brought the issue of vacuum polarization into the fray, it suggests
itself to ask whether other later contexts for vacuum polarization led to
similar surface proportionality. This is indeed the case for the
\textit{localization entropy}

This includes the identification of the (if possible most general) structural
properties which lead to broken symmetries in distinction to no symmetries. In
other words one is looking for an intrinsic mechanism which allows to
recognize the vitual presence of an original symmetry. There are two such
situations in QFT, the mentioned Goldstone spontaneous symmetry breaking,
whose signal is the appearance of a massless boson, and the Schwinger-Higgs
\cite{Schw}\cite{Hi} \textit{screening mechanism}, which typically leads to a
mass gap in gauge theories (independently discovered by Brout and Englert
\cite{Br-En}). As in the case of the QFT Goldstone spontaneous symmetry
breaking versus the solid state physics Heisenberg ferromagnet, it was
preceded by Anderson's \cite{An} discovery of an analog mechanism in condensed
matter physics.

In the intrinsic setting of QFT, the Goldstone theorem states that a conserved
current in QFT may not lead to a global charge as a result of bad infrared
behavior of some of its matrix elements; as mentioned before in order for this
to happen there must exist a "Goldstone boson" in the model i.e. a zero mass
particle which couples to the conserved current in a specific way in order to
prevent the large-distance convergence of the integrated current to the "would
be" charge. Kastler, Robinson and Swieca \cite{KRS} proved that a necessary
structural requirement in any covariant local QFT for this to happen is that
the spectrum reaches down to zero. By using the Jost-Lehmann-Dyson
representation Ezawa and Swieca \cite{ES} succeeded to sharpen this statement
by proving the existence of a zero mass particle which couples in a specific
way to the current. With this result the Goldstone theorem changed from a
statement about certain Lagrangian models to a structural theorem in QFT. The
insight gained into QFT was then transferred back by Swieca to solid state
physics in order to understand the connection of range of forces and broken
symmetries \cite{solid}.

The whole complex of conserved currents, including some subtleties in the
unbroken case caused by the ubiquitous presence of vacuum polarization clouds,
was nicely presented by Swieca 1967 in his Cargese lectures. Even after four
decades these notes \cite{Car} are still recommendable. This work on
spontaneous symmetry breaking brought Swieca the respectable Brazilian
\textit{Santista prize}. The quest for a profound structural understanding of
spontaneous symmetry breaking (as well as numerous attempts to exemplify
spontaneous symmetry breaking in concrete models) remained an area of living
research up to this day since it is of interest to explore the Goldstone
mechanism under the most general physical assumptions.

\section{The Schwinger-Higgs screening mechanism and the standard model}

The second way of breaking a symmetry, namely the Schwinger-Higgs mechanism,
is strictly speaking a a process of \textit{screening electric charges}. In
the formulation with pointlike covariant vector potentials and BRST ghosts it
is often called "gauge symmetry breaking" (see below). The charge screening
problem is not related the opposite problem namely to a large distance
divergence from integrating over zero components of conserved currents, but
rather to the question under what circumstances such integrals vanish. In that
case the conservation law of charges become ineffective and copious particle
production of "screened" particles would violate the charge selection rules
which holds in the electrically charged phase . \ Of special physical
interests for the discussion of screened charges are identically conserved
currents of the Maxwell type%
\begin{equation}
j^{\mu}=\partial_{\nu}F^{\mu\nu}\label{max}%
\end{equation}
Swieca showed \cite{scree} that the presence of a corresponding nontrivial
charge implies the existence of photons as well as a certain nonlocality of
the charge carriers with respect to the $F_{\mu\nu}$ observables resulting in
a weakened smoothness/analyticity properties of the electromagnetic
formfactors. The other side of the medal is the statement that a massive
"photon", which requires more analyticity, is only possible in case of a
vanishing charge (see also \cite{Bu-Fr}). In a QED-like theory with a would-be
charged scalar field there exists a phase in which this scalar field
contributes to its own screening and the resulting physical particle is not
subject to the charge superselection rules while the "photon" has turned into
massive vectormeson, in short one arrives at the \textit{Higgs
mechanism\footnote{Despite the similarities in the Lagrangians the point of
view in the paper by Higgs \cite{Hi} and similar publications by Kibble
\cite{Ki} as well as Brout and Englert \cite{Br-En} are quite different from
the present screening setting.}}. In the presence of electrically charged
spinor matter the scalar screening only affects the "Maxwell-charge" whereas
the global spinor charge and the related superselection rule continue to be valid.

Swieca was not only familiar with Schwinger's idea that QED may possess
another \textit{massive photon phase} (which goes back to the end of the 50s),
but he also contributed together with John Lowenstein \cite{Lo-Sw} some
beautiful work on a concrete two-dimensional model which Schwinger \cite{Schw}
had proposed in order to illustrate his idea of a massive phase in QED-like
gauge theories. In contrast to the Goldstone situation in which, according to
a well-known early argument in condensed matter physics \cite{Me-Wa},
spontaneous symmetry-breaking of a continuous symmetry group cannot occur for
d=1+1\footnote{In QFT this can be directly seen from the infrared-behavior of
the zero mass two-point function.}, there is no such dimensional restriction
for the Schwinger-Higgs screening mechanism and therefore Schwinger's model of
massless two-dimensional "QED" is a valid demonstration and also a reminder
that the mass-generating Schwinger-Higgs mechanism strictly speaking does not
deal with symmetry breaking. 

Since this screening mechanism has been discovered in the context of local
gauge theories, it is customary but somewhat misleading to call it (not in
Swieca's work) broken "gauge symmetry". To the extent that this refers to
local gauge invariance this may cause misunderstandings since the terminology
ignores the fact that the local gauge freedom in QFT parametrizes the
\textit{liberty of changing spurious ghost degrees of freedom} which leave no
trace in the physical cohomology space. It is however a valid physical
terminology inasmuch as it refers to the the symmetry associated with the
Maxwell charge which, as a result of screening, looses its superselecting
power with a resulting reduction of symmetry.

Neither Schwinger's nor Swieca's understanding of the mechanism of screening
was without historical precedent. In the setting of a quantum mechanical
Coulomb gas the idea goes back to Debeye (Debeye screening) \cite{Br-Fe}. The
point is that under certain circumstances the potential for large distances is
not of Coulomb but rather of Yukawa type; the quantum mechanical system
becomes self-screening. Screened states are closely related to plasma states.

In QFT charge screening does not need the actual presence of many Coulomb
charges, rather the vacuum polarization inherits this role. The result is a
much more radical kind of screening in which the unscreened and the screened
system are unitarily inequivalent and live in different Hilbert
spaces\footnote{The fact that the sceened Schwinger model in the limit of
short distances passes to the charged Jordan model illustrates this point
\cite{Jor}.}

It is somewhat ironic that the Schwinger-Higgs screening mechanism, whose
precise understanding is of crucial importance for contemporary particle
physics, is not as well understood as Goldstone's spontaneous symmetry
breaking with which it is sometimes confused ("fattening of the photon from
swallowing half of degrees of freedom of the massive complex field" so that
only a charge neutral real massive field remains). But, as pointed out. the
similarity of the Goldstone mechanism and that of the Schwinger-Higgs
screening does not go beyond formal Lagrangian manipulations. A more intrinsic
setting reveals that both kinds of symmetry breaking are very different. In
spontaneous symmetry breaking the integral over the charge density
\textit{diverges}, whereas in the screening case it \textit{vanishes}
\begin{equation}
\lim_{R\rightarrow\infty}Q_{R,\Delta R}^{spon}\psi=\infty,~\lim_{R\rightarrow
\infty}Q_{R,\Delta R}^{sreen}\psi=0
\end{equation}
A zero total charge is not a basis for a charge symmetry, i.e. the scalar
particles in the screening phase can be copiously produced. The physical
manifestation of the spontaneous breaking is the appearance of a zero mass
particle ("Goldstone boson") whose model independent existence, as mentioned
in the previous section, was established in a theorem (using the
Jost-Lehmann-Dyson spectral representation) by Ezawa and Swieca \cite{ES}
whereas the screening theorem showing the existence of a "massive photon" (at
the cost of loosing half the degrees of freedom of the complex scalar field)
is due to Swieca \cite{scree}. Note that the symmetry-breaking in the
screening mode is that of the breaking of the electric charge
conservation\footnote{It is important to remember that scalar QED has one
parameter more (the $g\left\vert \Phi\right\vert ^{4}$ term) than its spinor
counterpart. So the Lagrangian used by Higgs before adapting it to the
screened phase is identical to scalar QED, i.e. the ingredients of the Mexican
hat potential are already part of scalar QED.}. The issue of screening in
gauge theory was later taken up in \cite{Bu-Fr} were some of the arguments
used in Swieca's paper received additional mathematical support.

The standard version of the Higgs mechanism does not mention the screening
point of view. This does not render neither the Higgs-Kibble nor the
Brout-Englert presentation incorrect because at the end it is the correctness
of the renormalized correlation functions of the local observables and not the
physical ideas and mnemonic crutches which are used during their constructions
which defines its intrinsic physical content. 

The idea that scalar electrodynamics does not need any additional Higgs
particles nor a "Mexican hat potential", because everything to activate the
screening mode of QED is already there, seems to belong to the lost ideas
mentioned in the introduction. 

In the screening picture half of the complex scalar degree of freedom serve to
convert the photon into a massive vectormeson and the remaining real field
$R(x)$ has lost all symmetries, even R --%
$>$
-R.

To distinguish one viewpoint over others one must show that it explains facts
whose understanding in the other is not possible or unnatural. Since QFT is
founded on causal locality, every calculational device must be related to
problems of localization. In the next section it will be shown that the
modular localization distinguishes the screening viewpoint.

The relation between smooth momentum behavior and localization properties
which Swieca observed in the course of proving his theorem was the starting
point of Buchholz and Fredenhagen \cite{Bu-Fr2} who derived localization
properties from the gap hypothesis of the energy-momentum spectrum. They
showed that the worst localization which can happen in a theory which has a
mass gap and a pointlike generated local observable subalgebra is that one
needs semiinfinite spacelike stringlike localized fields to generate the full
algebra. In other words there is no use for generators which are localized on
higher than one-dimensional submanifolds.

In his proof Swieca noticed that in theories with a Maxwell structure
(\ref{max}) the nonvanishing of the charge requires the presence of certain
nonlocal properties which are absent in case of screening. In the 60s and 70s
there was the vague conjecture that electrically charged particles cannot be
particles in the sense of Wigner i.e. affiliated with irreducible
representations of the Poincare group. With other words there was a suspicion
that behind the infrared divergencies\footnote{The perturbative logarithmic
divergencies sum up to zero which is also what one obtains by direct
(nonperturbative) application of the large time LSZ limits.} of LSZ scattering
theory applied to QED there was something more dramatic than the infrared
treatment of Bloch and Nordsiek \cite{Bl-No} and the more QFT compatible
description of Yennie, Frautschi and Suura \cite{Y-F-S} revealed which led to
finite soft photon inclusive cross sections. There were some soluable two
dimension models in which the particle mass shell figuratively speaking is
sucked into the continuum so that instead of a particle the theory described
massive "infraparticles". In these models the fields which led to such
two-point functions were not pointlike but rather semiinfinite stringlike. But
it took another two decades to show that this is precisely what happens with
fields of electrically charged particles \cite{Buch}. Here the charge flux
through arbitrarily large surfaces (the quantum Gauss flux) assures the best
(tightest) possible localization cannot be better than a semiinfinite
spacelike string. This is quite interesting since the appearance of
necessarily noncompact localized objects in a theory which was thought to have
pointlike generating fields is somewhat unexpected. Swieca's observation
concerning bad analytic properties of electric formfactors in charged states
as opposed to the good analytic behavior in screened states is a reflection of
the different localization properties.

The string localization creates problems to deal with electrically charged
fields. In fact one representation is probably known to most readers. It is a
formal expression for a physical electrically charged scalar field%
\begin{equation}
\Phi(x;e)=~"\phi(x)e^{\int_{0}^{\infty}ie_{el}A^{\mu}(x+\lambda e)e_{\mu
}d\lambda}" \label{DJM}%
\end{equation}
This Jordan-Dirac-Mandelstam (DJM) expression appeared (according to the best
of my knowledge) for the first time in a 1935 paper by Pascual Jordan as the
best description for the field of a charged particle in a situation in which
non of the formally charged pointlike objects are physical. What makes the
situation computationally unwieldy is that this physical object has to be
introduced by hand, it is not part of the renormalization setting but its
perturbative construction requires constructing a separate formalism
\cite{Stein}.

On the other hand the screening situation leads to a different problem. As
with all situations involving massive vectormesons (including massive QED
\cite{Lo-Sc}) the gauge formalism (Gupta-Bleuler or BRST) has no intrinsic
physical meaning, its only purpose is that of a "technical catalyzer": it
helps to get over the power counting barrier at the expense of violating
quantum principles; once the renormalization has been done one can return to
the ghostfree pointlike vectorfields of short distance dimension $sdd=2$ times
logarithmic corrections. As in real life one is forced to reach a legal (gauge
invariant) result by illegal means and at the end the result justifies the
means. But in doing this, there remains a bad taste because one is forced to
move between two description which have no precise mathematical relation with
each other, namely the so-called renormalizable and the unitary gauge. Even
the best treatment of screening \cite{Stein2}, in which the physical content
of a massive vectormeson interacting with a selfconjugate (the charge has been
screened) scalar field is presented, has to struggle with this conceptually
unclear "switching problem" (which first was noticed first in massive QED
\cite{Lo-Sc}).

Behind the Schwinger-Higgs screening mechanism hides a fundamental problem
whose understanding is of importance for the future of the standard model and
more generally of interactions involving massive higher spin fields, namely
does the requirement of compact localization in the presence of interacting
$s\geq1$ require the presence of lower spin "satellites"? This would be
reminiscent of supersymmetry, except that in this case it would be related to
the most basic principle of QFT and not to a mind game of some physicists. A
better comparison may be the zero mass Goldstone boson which the theory needs
in order to break global charge conservation in the presence of a conserved current.

Whereas in Schwinger's original treatment it was very hard to identify the
gauge invariant content of the Schwinger model, the Lowenstein-Swieca
\cite{Lo-Sw} presentation clarified the chiral symmetry breaking and the
ensuing emergence of a $\Theta$-angle as a consequence of the Schwinger-Higgs
mechanism. In this way it became obvious that the gauge invariant content of
the model was generated by a free massive field and thus the physical content
became elegantly separated from gauge dependent unphysical aspects of the
Lagrangian setting in which Schwinger first presented the model. Among all
free fields, a massive field in two dimensions is very peculiar since its
short distance zero mass limit (as a result of its infrared property) defines
an algebra which has continuously many "liberated" charge sectors (so that the
massive model may be considered as a charge-screened version) with the charge
carrying operators being string-localized (localized on a semiinfinite
lightray). This has a vague analogy with the way quarks become "visible" in
the short distance limit of QCD. The gauge-independent intrinsic content of
the Schwinger model which consists in a two-dimensional neutral massive scalar
free field is capable to explain why for short distances the charge screening
passes to charge liberation\footnote{It is somehow easier to associate the
Schwinger model with the process of short distance charge liberation than to
start with free charges and go the opposite way of screening.}. There remains
however an important difference between the screening of charges, a process in
which the gauge potentials become associated with massive "photons", and
confinement of (generally nonabelian) charges, in which the charges associated
with representations of the fundamental theory are "confined" and only their
composites appear in the physical spectrum of the theory. Swieca and
collaborators have made attempts to explain the difference between screening
and charge confinement in a mathematically controllable two-dimensional
context \cite{K-S-S}\cite{B-R-S-S}. There are however limits to analogies for
screening versus confinement concepts in higher spacetime dimensions. In d=1+1
all the models used for that purpose were superrenormalizable and hence they
fulfilled the requirement of asymptotic freedom in an almost trivial manner;
for strictly renormalizable theories this is a somewhat harder problem, even
if they are two-dimensional. 

In 4-dim. QCD it took the computational ingenuity of Politzer, Gross and
Wilszeck to arrive at the consistency check for the asymptotic freedom
conjecture. If the model is soluble, as the strictly renormalizable
factorizing Gross-Neveu model, one is able to rewrite the Callan-Symanzik
parametric differential equations in terms of physical mass parameters from
where one can read off a proof of asymptotic freedom. In QCD one does not know
how to arrive at a physical reparametrization; this is of course related to
the lack of knowledge about the physical confinement phase. A full proof
beyond a consistency argument is probably not possible without knowing more
about the confinement problem.

Nowadays it is hard to imagine that at the time of Swieca there was still
resistance against the Schwinger-Higgs screening mechanism. Swieca once told
me that he was not able to convince Peierls that a massive phase of gauge
theory could exist; Peierls apparently insisted that the quantized Maxwell
structure cannot be reconciled with massive photons.

Swieca's work on charge screening and the mass spectrum was deepened by
Buchholz and Fredenhagen \cite{Bu-Fr}  who succeeded to supply it with the
mathematical rigor and the conceptual astuteness of local quantum field
theory. The weak point in Swieca's screening proof was related to certain
analytic properties in particle momenta of formfactors. Buchholz and
Fredenhagen proved these properties and realized that they can be used to
settle other even more ambitious problems. In fact this sent these authors on
a much more general track of investigating the connection between localization
and particle spectra . Their physical motivation was to reconcile the
non-abelian gauge structure with the massiveness of the QCD particles. The
main result of this work (which considerably widens the realm of QFT) in
modern parlance says that assuming the existence of (pointlike) local
observables and the existence of a spectral gap (expected in QCD as the result
of confinement), the generator of charges are covariant semi-infinite
space-like string fields \U{3a8}(x, e) where the unit vector e represents the
space-like direction of the semi-infinite string which starts at $x$; in
particular there is never any need to introduce generating quantum fields into
QFT with a mass gap whose localization goes beyond point- and string-like
extension (be aware this is not string theory!). All objects with larger
localization can be obtained from interacting string-like fields. Point-like
fields constitute a special case when the field is $e$-independent. The
methods of algebraic QFT used by those authors are not model-specific and it
is up to now an open problem to give an intrinsic physical characterization of
what is meant by a non-Abelian Maxwell structure. So what the authors ended up
with was a framework allowing semi-infinite string-localized fields to arise
from rather general assumptions about the energy-momentum spectrum but it is
presently not possible to decide whether this mechanism is taking place in
QCD. In any case this illustrates in a nice way that the legacy of an idea may
sometimes pass through methodological improvements from one problem to another.

The history o\cite{Bu-Fr2}f conserved currents, which led to two very
different kind of symmetry breakings illustrates in an interesting way how the
legacy of an idea passes through methodological improvements from one problem
to another. In the next section it will be argued that this subject continues
to play a dominant role in present research.

\section{The unfinished business of gauge theory}

A better understanding of the physics behind gauge theories requires a basic
conceptual revision of local gauge invariance in terms of a more intrinsic
description of interactions involving $s\geqq1$ fields. If these fields are
massive, the covariant description of the (m%
$>$%
0,s=1) Wigner representation with the smallest short distance dimension (sdd)
is in terms of a vector field with $\partial^{\mu}A_{\mu}(x)=0$ with $sdd=2$.
This value is above the power-counting limit of renormalization theory which
is $sdd=1$ Bosons and $sdd=3/2$ for Fermions. The minimal sdd increase with
spin. But by allowing massive covariant generating fields which are
semiinfinite string localized, one can always reach sdd =1 \cite{MSY} which
then opens the possibility to construct interactions which fulfill the power
counting criterion and are therefore candidates for renormalizable models.

In the massless case there is a much more compelling reason for stringlike
"potentials" instead of pointlike "field strengths". This and the reason for
the quotation marks becomes clear if one compares the possibilities for
pointlike covariant fields in both cases.

Whereas in the massive case the infinitely many covariant dotted/undotted
spinorial field associated with a unique Wigner representation $(m>0,s)$ obey
the following inequalities between the spinorial indices and the physical spin
\cite{interface}%
\begin{align}
&  \Psi_{m>0,s}^{(A,\dot{B})},\ \left\vert A-\dot{B}\right\vert \leq
s\leq\left\vert A+\dot{B}\right\vert \label{spinorial}\\
&  \Psi_{m=0,s}^{(A,\dot{B})},~s=\left\vert A-\dot{B}\right\vert
,~~~s~helicity~\nonumber
\end{align}
the Wigner representation theory severely limits the spinorial pair for a
given helicity in the massless case to the equation in the second line.
Different from the situation in classical field theory, the intrinsic Wigner
representation theory is not consistent with certain covariant pointlike
fields which are allowed for massive representations. This is not the fault of
the unitary Wigner representation but rather the result of a basic clash
between the positivity (Hilbert space) of quantum theory and the pointlike
localization properties of covariant fields.

In the intrinsic (not relying on classical quantization parallelism) massless
Wigner representation theory and also in QED (and the gluons in QCD) there is
no place for pointlike vectorpotentials for which $s=1$ and $\left\vert
A-\dot{B}\right\vert =0$. On can of course enforce the existence of pointlike
covariant objects by leaving the setting of quantum theory and playing the
ghost game known under the acronym of Gupta-Bleuler or BRST (which has a
larger region of applicability than Gupta-Bleuler). There are certain formal
tricks leading to a subalgebra of genuine local quantum observables in an
appropriately constructed Hilbert space, which can be done with such a
formalism; but there are also serious limitations especially if it comes to
the description of any physical object which cannot be pointlike generated as
the electrically charged fields.

The big surprise (at least to me) is that the full spinorial formalism for
massless fields i.e. \textit{the full spectrum of spinorial indices} in the
first line (\ref{spinorial}) can be recuperated if one admits semiinfinite
spacelike strings (any confusion with string theory must be
avoided\footnote{The objects of string theory are not string-localized in any
material i.e. non metaphorical sense \cite{ST}\cite{foun}.}) i.e. fields of
the form $\Psi_{m=0,s}^{(A,\dot{B})}(x,e)$ where $e$ is a spacelike string
direction; as in the massive case their sdd is below the power-counting
threshold. The prize to pay is a weaker localization, namely semiinfinite
stringlike instead of pointlike.\ This requires new physical ideas of how to
cope with such string-localized fields in renormalized perturbation theory
(extension of the Epstein-Glaser procedure) instead of finding formal tricks
of how to handle ghosts in order to reach the saving shore of quantum theory.
Sharing with Swieca Haag's local quantum physics conceptual setting of QFT. I
firmely believe that in all cases of use of indefinite metric in problems of
QT it enters as a placeholder for a deep problem which still has to be solved.

A $s=1$ vector potential is of the form $A_{\mu}(x,e)$ with the same momentum
space creation/annihilation operators as in the field strength but with
different intertwiners $u(p,e)$ from the Wigner to the covariant
representation. Similarly a $s=2$ potential associated with a field strength
(which has a linearized form of the Riemann tensor) would be of the form of a
symmetric tensor $g_{\mu\nu}(x,e).$ These stringlike free fields would
fluctuate in $x$ \textit{and} $e,$ In the case of the vectorpotential these
fluctuations of the spacelike unit vector $e$ in 3-dimensional de Sitter space
(in order to highlight that the infrared directional fluctuations are
indistinguishable from pointlike ultraviolet fluctuations) relieve the
fluctuations in $x$ such that short distance dimension in $x$ is reduceds
namely $sdd_{x}A_{\mu}(x,e)=1.$

In the massive case there is no \textit{structural} (representation
theoretical) \textit{necessity} to introduce string localized potentials, but
they nevertheless exist and come with the attractive property of $sdd_{x}=1$
independent of spin which at least potentially increases the existence of new
renormalizable interactions involving stringlike objects. If in case of the
Schwinger-Higgs screening model the string-localized formulation fulfills also
the finer points of renormalization theory as generalization of the E-G
iteration to string localized fields (for which there are good indications),
this would lead to a unified treatment without the metaphoric aftertaste from
being forced to move between two different descriptions \cite{charge}.

The new formalism\footnote{There is no linear pointlike generated subspace
generated by applying the interacting gluon potential to the vacuum in the
Yang-Mills case.} is expected to address those problems which remained outside
the range of the gauge formalism under the new heading \textit{how to deal
with nonlocal physical objects i.e. objects which cannot be described in terms
of pointlike generating fields}. They would be generated by semiinfinite
strings, as the example of electrically charged fields and their associated
\textit{infraparticles} \cite{infra} show. In Yang-Mills theories one expects
the existence of a much stronger form of semiinfinite string localization
which may be the key for the understanding of the expected "invisibility" of
gluons and quarks i.e. the intended meaning of "confinement". Whereas in QM
particles can be confined into a compact cage by a suitably chosen potential,
the only resource which is available to QFT is the noncompact localization of
its most basic constituents as a necessary prerequisite for not entering
counters which register compact localized objects. However string-like
localization alone is not enough, as the very visible string-like localized
charged particles of QED demonstrate. One must show that the interacting gluon
strings create states from the vacuum which are of a very different kind
\cite{charge}, a point to which we will return later on.

Wigner's representation theory does not envisage a necessity to introduce
generating wave functions (and associated fields) localized on higher than
one-dimensional subspaces. Although the problem of interacting operator
algebras has more possibilities for trans-pointlike localizations, there are
rather convincing arguments that even there one never has to go beyond
semiinfinite string-like localized generators \cite{Bu-Fr2}. In particular the
principles of QFT do not support the existence of models in which the
generators are "brane-localized", although generators of holographic
projections and of wedge-localized algebras may play useful role in certain constructions.

Since the higher spin string-localized potentials are not Lagrangian objects,
the standard perturbation setting (either in operator or in functional
integral form) is not an option. Hence the main new technical problem is the
adaptation of the locality-based perturbative Epstein-Glaser iteration to
string fields. The first test of the new theory should be the construction of
the stringlike charged fields by keeping all the points $e$ in 3-dim. de
Sitter space of the vector-potential $A_{\mu}(x,e)$ different during the
computation and taking their local limit only at the end, just like in the
construction of composite fields from Wightman functions. Note that the reason
for introducing a ghost formalism ala Gupta-Bleuler or BRST is dispensed with
since the fields already have their lowest operator dimension and the BRST
invariance is nothing but $e$-independence as the characterization of a
subalgebra, the gauge-invariant algebra of the gauge approach. Since in the
zero mass s=1 Wigner case the $A_{\mu}(x,e)$ exists and has the minimally
possible dimension $sdd_{x}=1$, its existence does not have to be bought by
sacrifying the Hilbert space of quantum theory via the (intermediary) presence
of ghosts. This has an extension to any zero mass $s>1,$ there always exists a
"potential" with $ssd_{x}=1$ whose appropriate (higher) derivative belongs to
the admitted pointlike "field strength" according to the second line in
(\ref{spinorial}).

It is worthwhile to emphasize again that in the massive case the necessity for
either ghosts or string-localization enters through the back door: it renders
a theory whose higher spin fields have increasing short distance dimensions to
lower dimensional "potentials" with $sdd_{x}=1$ which lead to renormalizable
(in sense of power counting) couplings. Whether the step of working with
string-localized potentials is just a "catalyzer" to overcome the
renormalization barrier, or whether there remain genuine string-localized
objects in the resulting interacting theory is not clear, an educated guess
would favor that at least in screening modes the resulting theory has a
pointlike description.

In QED the electrically charged objects are string-localized (infraparticles)
and the physically important currents and field strengths are point-localized;
the string-localization of the vector-potential has been fully transferred to
the electrical charge carrying field, the string-localization of the potential
itself remains harmless; with the possible exception of Aharonov-Bohm like
effects (and their generalizations to higher s) there are no \textit{direct}
consequences since the field strength which remains point-like.

In the gauge setting it is difficult to pinpoint the mechanism which prevents
charge generators to be pointlike and forces them to be stringlike. Each ghost
formalism is totally pointlike, but this has nothing to do with the physical
content; it only serves to save the recipes of the standard perturbative
renormalization formalism at the expense of the physical meaning of locality.
If one sacrifices the most important principle of QFT which is locality (and
together with it the Hilbert space, which is the holy Grail of QT), one should
not be surprised that one misses out on the localization of charged objects.

Vacuum expectation values of charged fields are charge neutral. In that case
one expects that one can arrange the semiinfinite strings in such a way that
the infinitely extended parts compensate and only finite "gauge bridges"
between opposite charged points (the starts of the semiinfinite strings)
remain\footnote{Only neutral pointlike composites of charged fields are free
of strings and hence are associated to the pointlike generated observable
subalgebra.}. However this does not mean that one has gotten rid of the
infrared aspects of the semiinfinite strings. When one tries to extract
particles via infinite timelike limits, the infinitely extended strings
return. There radically different infrared nature explains why not only
scattering theoy is conceptually very different but also why even the
one-particle states suffer a drastic conceptual change and become
"infraparticles". Different from the logic in the standard treatment of the
infrared problem which is done without paying attention to spacetime
properties \cite{Bl-No}\cite{Y-F-S}, the delocalization of charged particles
to infraparticles via the mechanism of localization-caused vacuum
polarization\footnote{In the standard QFT situation the infinite vacuum
polarization cloud comes from infinite energies (the local sharpness of a
"bang" onto the vacuum) whereas for a bang with an electrically charged field
there are also infinitely many infrared photons in the charged bang state.} is
the primary mechanism and the breakdown of scattering theory and its
replacement by a theory of cross sections with finite infrared photon
resolution is a consequence.

This breakdown of the scattering theory via the infra-particle mechanism is
much more radical than that resulting from quantum mechanical Coulomb
scattering which leads to the appearance of logarithic phase factors which
present the large time convergence of amplitudes and requires to formulate
asymptotic convergence in terms of cross sections \cite{Dollard}. In that case
(as in all cases of QM) the structure of one particle states is not affected,
whereas in QED the free irreducible one electron Wigner state becomes a
reducible infraparticle state whose free mass shell has been dissolved into
the continuum and who carries a permanent photon cloud within an energy
resolution which can be made arbitrarily small but cannot vanish. The fact
that the lattice description has problems to describe these properties even in
the abelian case cast some doubts on the believe that electrically charged
fields and particles and their nonabelian counterparts can be described in
lattice setting.

A nice discussion of this problem in terms of a breakdown of n-fold localized
tensor product states states can be found in \cite{Haag}\cite{Enss}. In fact
the string direction of a charged particle spontaneously breaks the Lorentz
covariance because a physical unit charge state contains actually infinitely
many helicity s=1 carrying infrared photons. This leads us back to Swieca's
observation about the lack of expected analyticity in formfactors in case the
formfactor is taken in charged states. The reason for this is the noncompact
localization of charged infraparticles.

In the screened scalar QED the only purpose of the stringlike vector-potential
is to render the coupling renormalizable; there is no delocalization on the
scalar matter field, the originally complex scalar field just looses half of
its degrees of freedom and becomes a real i.e. chargeless field \cite{Stein2}.
This suggests the following approach: select among all renormalizable (in
sense of power-counting) couplings between a string-localized massive
vectorpotential with a real field those for which for which the real field is
pointlike. In general there will be stringlike localized matter fields and
only certain composites of them will be pointlike. Such a situation is on hand
in Yang-Mills models; whereas generic polynomial interactions between
stringlike vectorpotentials are expected to lead to situations without
pointlike composites, Yang-Mills couplings between gluons lead to composites
which are pointlike (the "gluonium" field). The form of the renormalizable
interactions and the form of the local polynomials should be solely
determinded in terms of pointlike localization requirements (the locality
property overrides group theory); the gauge group is only a mnemonic devise
for finding the polynomials in terms of string localized gluons.

The new principle to look for pointlike polynomials within a set of
power-counting renormalizable couplings of string-localized objects could well
lead to a wealth of new models involving higher spin fields.

The reader who is familiar with the gauge theoretic formulation may have
noticed that the string vector potentials have a formal similarity with vector
potentials in the axial gauge $e_{\mu}A^{\mu}(x,e)=0=\partial_{\mu}A^{\mu
}(x,e)$. In fact the difference is mainly one of interpretation. From the
gauge theoretic point of view $e$ is a gauge parameter. But then the
occurrence of the severe perturbative infrared divergencies, which prevented
well-defined perturbation calculations in the axial gauge, remains a mystery;
why should an inert gauge parameter create such messy infrared problems? The
interpretation in terms of string-localization is not imposed but follows from
the form of the commutator. As a result one also the unitary transformation
law in which $e$ participates in the transformation law instead of staying
inert as in the axial gauge setting. The long range nature of string-localized
fields explains this (since $e$ is not a bona fide gauge parameter) and tells
one what to do; mathematically the vectorpotential is an operator-valued
distribution in $x$ and $e.$ In order to obtain the physical content of the
DJM formula (\ref{DJM}) for a charged field, not by hand but within a
perturbative formalism which includes string-localized fields, it is necessary
to keep all the inner $e^{\prime}s$ in loop graphs distinct and study their
confluence limit at the end. This problem of fusing points in the
3-dimensional de Sitter space of unit directions is expected to resemble the
formalism of constructing composites.

There remains of course the question why these observations were not already
made at the time of Swieca, in particular since they fit so perfectly into the
screening setting and the perceived nonlocality of electrically charged fields
at that time on the basis of the quantum Gauss law. It was only necessary to
notice that the gap in the spinorial description of the ($m=0,s\geq1$) Wigner
representations can be filled with string localized spinorial free fields
$\Psi^{(A,\dot{B})}(x,e).$ There is no easy answer to that question, but I
think the fact that the role of localization as the dominating physical
principle of QFT was not yet fully appreciated explains the missed chance to a
certain degree.

The change came in more recent times when the concept of modular localization
was discovered \cite{Bert}\cite{BGL}\cite{MSY}. Only then it was finally
possible to understand the field theoretic content of the third Wigner
representation family of massless infinite spin representations, the first
being the massive and the second the massless finite helicity representation
family. Generations of physicists, including Steven Weinberg, tried in vain to
force this rather large representation family into the scheme of pointlike
fields. The recognition that this third family has no compact localization
\cite{BGL} and can be described in terms of semiinfinite string localized
fields \cite{MSY}. This was the eye opener for looking at the mentioned
spinorial gap of the massless finite helicity representations.

This episode shows the importance of keeping unsolved or only partially solved
problems in one's memory even in the presence of the widespread opinion that
they have been solved (gauge theory, electrically charged fields,
Schwinger-Higgs screening) or rendered irrelevant in the shadow of a theory of
everything. The idea that one is in the possession of a new theory which
either explains the old problems or renders them superfluous is as
presumptuous now as it has been in the past. The underlying philosophy at the
time of Swieca was that it possible in particle theory to come to gradual
conceptual unification, but certainly not that of a theory of everything
(TOE). But also the idea that by playing with exclusively with "effective QFT"
one could come to new insights would have appeared strange to the QFT
community of which Swieca was a member. There are of course areas in physics
(solid state physics, quantum chemistry) in which there is no hope to derive
and describe observed phenomena from fundamental laws. The fascination with
the various gauge theories in the standard model comes from the conviction
that the basic problems of particle theory can be understood in terms of a few
fundamental laws.

But even in QED this is presently wishful thinking, not to mention its
nonabelian version. The important step is to make some conceptual headway in
the largely unknown landscape of QFT and then exemplify the new point in a
reasonably controlled model; this was the way progress came about in the 60s
and 70s. At no point did people think in those times that QFT is a theory
whose foundations are known apart from some details. Rather the prevalent
philosophical view was a certain astonishment that a more than 40 year old
theory had relinquished so few of its secrets. All other physical theories
offered illustrative mathematically controllable models in the presence of
interactions, not so QFT about which one only knew free fields and some
low-dimensional near free models without a nontrivial scattering matrix.
Fortunately this situation has improved somewhat (see next section) but it
still would be impudent to claim that QFT is a largely known subject. Even
some usually optimistic followers of the gauge principle became recently aware
to their surprise that gauge theory is not an intrinsic concept since by
looking at the \textit{physical} local fields of a gauge theory it is
impossible to decide whether the correlations come from a gauge theory or some
non-gauge description\textit{\footnote{Under the influence of studies of
\textit{duals} of gauge theories, the \textit{nonintrinsic nature} of the
concept \textit{of gauge theory,} which was a minority opinion at the time of
Swieca, seems to have been accepted by a majority of particle physicists.}}.
The description of localization of electrically charged fields and particles,
charge-screening and higher spin interactions, as well as the fate of gluon
and quark degrees of freedom are unsolved fundamental problems which
"effective" QFT does not address but whose solution it needs for its credibility.

The metaphoric ideas in an article about the legacy of Swieca on issues about
which most people have quite solidified (but not solid) notions may sound
provocative, but they are only reminders that most of the foundational
question have remained open. After 40 years of research there is the pressing
problem whether the present setting of gauge theory is the appropriate
framework. The modular localization concept re-interprets the gauge invariant
local observables as $e$- independent subobjects of a theory which includes
string-like generating fields. In doing this certain objects as electrically
charged fields and their associated infraparticles which as "nonlocal" objects
were hitherto outside the gauge formalism (and had to be defined by hand
(\ref{DJM})) are now incorporated. it becomes clear that these open problems
related to interactions with non-pointlike fields cannot be answered in the
present setting about gauge theory.

Some of these questions have been successfully addressed already by Swieca
who, in lack of modular string localization, connected the failure of
pointlike locality of charged states with weaker analytic and smoothness
properties of electric formfactors i.e. matrix elements of currents in charged
states. The problems commented on in this section are in my view extensions of
ideas which got lost, but whose continuation may be helpful to overcome the
present crisis of almost 40 years of stagnation on fundamental problems. After
40 years of partially successful dominance of "effective" QFT, it is time to
again turn the foundational wheel on renormalizable higher spin interactions.

One extremely important aspect which is very present in Swieca's research, but
has been almost completely lost in contemporary research, is the notion of
\textit{intrinsicness} of a property or prescription which was used in a
particular construction. For example in the Schwinger-Higgs screening phase of
scalar QED (often called the Higgs-Kibble model) extended by a coupling to a
Dirac spinor there are two ways of having a massive spinor, either by starting
with a massive spinor or by using a Yukawa couping of the (massless) spinor
and claiming that the Higgs mechanism created the spinor mass. The second way
is referred to in semi-popular publications as "the God particle creating the
masses of quantum matter". But there is no intrinsic meaning of such a
metaphoric language unless one exhibits a property which allows to distinguish
the two. Of course this also applies to the concept of charge screening. Only
after one has reduced a property of a model of QFT to realization of the
causal locality principle (as in the previously mentioned case of
infraparticles following from the quantum version of Gauss law) an autonomous
understanding has been reached.

In particular there is no intrinsic meaning to the terminology "gauge
principle". The word "gauge theory" does not refer to a physical principle,
but only to a particular computational tool to extract local observables in a
theory about which the remaining physical aspects are unknown. Whereas this
was always observed by physicists with a strong LQP background, it only begun
to be appreciated by people outside after it was observed that certain gauge
theories were dual to gauge theories with a different gauge group or to
theories which are not of the gauge type. The main message of this section is
that one only arrives at the intrinsic meaning of a property once one
succeeded to trace it back to localization aspects.

Under this caveat all terminologies in QFT should be occasionally
reconsidered. This includes also the Schwinger-Higgs screening. Whereas the
quantum mechanical Debeye screening or the plasma state can be viewed as a
particular phase of a quantum mechanical Coulomb system, the view of scalar
QED and the Higgs model as different phases of the same Lagrangian QFT is more
precarious because QFTs with a different particle spectrum are globally
inequivalent to each other and the insight that they are locally related is
often out of reach. But since QFT contains its intrinsic interpretations this
does not generate any harm as long as one remains aware about these problems.

\section{Factorizing models and Swieca's contribution to "nuclear democracy"}

Another interesting idea of Swieca, which I consider an important part of his
legacy, has to do with massive 2-dimensional factorizing models. Some
introductory remarks are necessary. This research goes back to certain
quasiclassical observations by Dashen, Hasslacher and Neveu \cite{DHN}
suggesting the integrability of a family of d=1+1 theories including the d=1+1
massive Sine-Gordon- and Thirring- models.

The first attempts to understand the particle spectrum in connection with the
S-matrix of these models pointed to the old S-matrix bootstrap approach and
led to a modest revival of the old bootstrap ideas, but now specialized to
factorizing 2-dimensional models \cite{STW}. This bootstrap program. which was
so exuberantly praised in the 60, and fell out of fashion after the discovery
of QCD, finally found an interesting (albeit more modest) explicit realization
in form of an infinitely large family of d=1+1 models with a factorizing but
nontrivial S-matrix fulfilling the crossing property \cite{KTTW}.

In addition it was found that if one abandons the ideology of S-matrix
supremacy over QFT, including the metaphorical hope that the S-matrix
bootstrap by some magics selects a unique TOE (theory of everything), and
instead considers the classification of factorizing S-matrices as the first
step in a construction of "factorizing QFTs", one ends up with an very rich
quantum field theoretic harvest\footnote{At this point other actors (the
Zamolodchikovs, Faddeev, Witten, Smirnov) entered who brought important knew
ideas. The present status of the bootstrap-formfactor program has many
contributers and its review is not the aim of these notes.} \cite{KW}. The
models confirm the \textit{nuclear democracy} idea which results from the
locality principle of interacting QFT, namely all particle states with the
same superselecting charge quantum number which have the same charge are
necessarily coupled with each other as illustrated for formfactors under the
heading of Murphy's law in the introduction. A special corollary of this tight
internal connectivity of states in QFT is the principle of nuclear democracy
for bound-states which says that different clusters of particles which share
the same superselected charge lead to the same bound states. The cluster may
already contain bound states hence bound states may be viewed as being
composed of their own kind. This nuclear democracy property removes the
standard hierarchical elementary-bound order which dominates QM; the only
remaining hierarchy is that of fundamental and fused charges. But even this
becomes blurred due to the existence of the "self-equivalence": Sine-Gordon
soliton (massive Thirring field) $\simeq$ Sine Gordon soliton + arbitrary
admixture of Sine Gordon stuff.

This even holds if the masses of the particles of the original particles go to
infinity \cite{KKS} and in this way become unobservable (confined "quarks"),
showing that there is no contradiction between nuclear democracy and this
simple kind of confinement.

In interacting theories locality does not permit the field states of the
infinitely many composite fields with the same charge to have vanishing mixed
two point function. The states of particles belonging to different
superselected charges are of course orthogonal, but those corresponding to
different composites with the same fused charge lead to the same particle
states, apart from the fact that their composite interpolating fields have to
be renormalized by different constants. In some sense this democracy principle
makes QFT conceptually simpler than QM; but it also creates immense
computational problems if one tries to use similar operator methods as in QM.
The path from the factorizing S-matrix to a uniquely associated QFT goes
through the construction of formfactors i.e. of multi-particle matrixelements
of operators.

Swieca's interest in this rich class of controllable models arose mainly from
the possibility to test certain general conjectured structural properties of
QFT which are outside the range of perturbation theory. He realized that
factorizing models presented a rich theoretical laboratory for testing ideas.
One such idea was his conjecture that the principle of nuclear democracy may
permit to define and construct certain models in a completely intrinsic way on
the basis of the "minimal" version of nuclear democracy, without referring to
a Lagrangian. For example his definition of a "minimal" factorizing Z(N) model
is that of a factorizing model of particles with N charges numbered as n=0, 1,
..., N-1. The vacuum belongs to n=0, n=1 represents the "fundamental" particle
whose N-fold composition leads back to the vacuum sector, so that its N-1 fold
composition must play the role of the anti-particle, the N-2 composite is the
antiparticle of the n=2 bound state etc. This "minimalistic" realization of
this "the antiparticle as a bound state of N-1 particle" principle led to a
unique S-matrix \cite{Ko-S} and more recently also the formfactors of this
Z(N) model have been constructed \cite{B-K} in this way. This recent result
also confirmed that the only consistent field statistics (field commutation
relations) which one can associate to this model is the abelian braid-group
statistics as postulated by Swieca.

Most of the factoring S-matrices leading to uniquely associated QFTs are
outside the Lagrangian framework\footnote{This is to be expected since the set
of factorizing S-matrices is nuch larger than what can be encoded into the
local renormalizable coupling of fields and since every factorizing unitary
crossing S-matrix has precisely one set of crossing formfactors and hence one
QFT.} and the Z(N) and the chiral SU(N) model are representative
illustrations. With the conceptual framework of the Haag "school" in the
background, Swieca belonged to the (at that time) minority of particle
physicists who believed that the Lagrangian quantization approach to QFT does
not exhaust the richness of QFTs; after all it is based on a strange
parallelism of the more fundamental QFT to its less fundamental classical
counterpart which is hardly tolerable frm a philosophical point of view. By
now one knows that only a small fraction of factorizing models are
"Lagrangian" and the Z(N) model was perhaps the first non-Lagrangian model.
This is so because the richness of factorizing unitary S-matrices with
crossing property is much larger than what can be encoded into local coupling
of fields.

The chiral SU(N) Gross Neveu model resembled the Z(N) model concerning the
minimalistic antiparticle description and anyonic statistics, but assigns an
additional problem which attracted Swieca's attention \cite{GN1}. This had to
do with the question of how the apparent chiral symmetry breaking could be
reconciled with the Mermin-Wagner theorem and its much simpler field-theoretic
analog (infrared behavior of the two-point function in d=1+1 \cite{Col}) which
forbids a spontaneous breaking of a continuous symmetry in two dimensions.
With the hindsight of abelian charge-creating infrared-clouds in two
dimensions from previous work, Swieca et al. \cite{GN2} proposed a symmetry
protecting (from the S-matrix point of view restoring) mechanism caused by
infrared clouds\footnote{The Coleman theorem is not mention in the paper but
its knowledge is not of much help for figuring out the concrete restoration
mechanism in the model at hand. The existence of two different proposals from
just knowing the S-matrix demonstrates this.}. This was a different mechanism
from that proposed by Witten \cite{Witten} in the same model for the same
reason. Witten's proposal was further elaborated by Abdalla, Berg and Weisz
\cite{ABW}. But on the pure S-matrix level it was not possible to decide which
off-shell version was correct. In a recent paper \cite{BFK}, the formfactors
of this model have been constructed and their result clearly selects the
solution of Swieca et al.

The plethora of two-dimensional commutation structures led to the question
whether for those two-dimensional models which described the scattering of
particles, the statistics in the sense of field commutations is already
reflected in the one-particle states. This is certainly the case in higher
dimensional QFT. The answer was negative, i.e. two-dimensional particles are
statistical "schizons" since the fields associated with the particle can
always be changed by multiplying it with a disorder variable \cite{schiz}.
Since the statistics is related to crossing, the bootstrap-formfactor
construction of factorizing models selects a particular assignement which, if
desired, may be changed after the theory has been constructed. \ According to
the spin-statistics theorem this is not possible in higher dimensions. In
d=1+2 QFT the (braid-group) statistics is determined in terms of the
(anomalous) spin and this connection is already pre-empted in the setting of
Wigner's classification of one-particle states \cite{Mu}. The statistics in
the sense of field commutation relations is also intrinsic in d=1+1
\textit{conformal} theories.

Swieca's main interest was focussed on constructive aspects of QFT (in
particular the use of \textit{low-dimensional controllable models as a
theoretical laboratory}\footnote{In his own words \cite{Fort}:
\textit{Two-dimensional spacetime, despite all its peculiarities has proved
many times to be a fruitfull theoretical laboratory where one can test a
number of ideas in soluable models and many times draw inspirations for more
realistic models}.}) No-Go theorems did not belong to his range of interests.
But on one occasion, when he was convinced that an interesting proposal would
not stand up to physical requirements of macro-causality, he also proved a
No-Go theorem \cite{M-S}; the object of the critique was the Lee-Wick proposal
of using complex (+ complex conjugate) poles in Feynman propagators. Together
with one of his students he showed that by reformulating the problem into a
Yang-Feldman setting, the use of indefinite metric can be avoided and the
problem with causality appear in sharper focus. It turns out that the Lee-Wick
mechanism is untenable since it even violates the crudest form of macro-causality.

This No-Go statement should be viewed in the context of a long list of failed
attempts to maintain Poincar\'{e} invariance without micro-causality
\cite{Sch}. In recent times the nonlocality aspect reappeared in the veil of
"noncommutativity" through the backreaction of string theory on QFT. Since the
hallmark of quantum physics as opposed to classical physics has been
noncommutativity, this terminology needs an explanation. Noncommutativity in
the contemporary context means imposing a noncommuting structure on euclidean
functional integrals or modifying the real time formulation directly so that
the spacelike commutativity is violated. The construction of noncommutative
theories is a special way to obtain non-local theories. Apart from attempts
being guided by ideas from quantum gravity (absence of small black holes whose
presence would make any measurement impossible), most of the proposals suffer
from the lack of conceptual reasoning which, as a result of use of
sophisticated mathematics, is often not visible to the untrained eye.

This becomes especially evident if one compares the conceptual level of
present understanding with that during the two decades 60-80. In those days
the notion of causal locality played a central role in the interpretation of
QFT and it was generally acknowledged that the physics of momentum space (e.g.
Feynman rules) has to be derived from localization of states and locality of
operators. The Fourier transform of a translationally covariant operator has
apriori nothing to do with the energy-momentum of an object registered in a
counter, rather it is the mass-shell momentum in the sense of a geometric
relation between two asymptotically timelike removed events which lead to a
physical interpretation to the momentum space. It was generally accepted that
even if one is forced one day by new experiments to abandon micro-causality,
there is a minimal set of macro-causal requirements which are indispensable
for any kind of particle physics i.e. which are the properties one must keep
in any kind of relativistic particle theory. According to considerations going
back to Stueckelnberg, the causal rescattering (in QFT often referred to as
the \textit{causal one-particle structure}) insures the absence of timelike
precursors and together with the cluster property of the S-matrix constitute
the time- and space- like aspects of macro-causality. Although it was clear
that the Lee-Wick proposal violated micro-causality, the violation of
macro-causality and hence its physical inconsistency only became exposed in
\cite{M-S}.

The problem of whether one can weaken microcausality in a physically
consistent way has remained in the forefront after Swieca's death in December
of 1980; although the motivations for exploring non-local theories have been
changing. Newcomers to QFT notice over short or long that locality is an
extremely restrictive requirement, but it is much harder for them to realize
that the conceptual problems of physically acceptable relativistic nonlocal
theories are even more severe; the idea of constructing consistent models
which are just "a little bit non-local" has been one of the most treacherous
since the late 50s when physicists became interested in this topic.
Poincar\'{e} covariance and energy positivity severely limit such a spatial
fall-off of the commutator (for a review of attempts at non-locality
\cite{Sch}). For example the commutator cannot decay faster than the Yukawa
exponential if one wants to prevent falling back at a local theory. The only
non-local setting which is under mathematical control and fulfills all
macrocausality requirements which one is able to formulate in terms of
particles, is the direct particle interaction scheme of Coester and Polyzou
\cite{Co-Po}. But this has not and cannot be obtained by modifying the
construction of QFT since the way in which the cluster property arise in these
quantum mechanical models is not compatible with second quantization (but
nevertheless correct).

Physicists before the 80s had to learn the hard way about the conceptual
barriers which one confronts in departing from the realm of locality. Looking
at the lighthearted manner in which formal games for getting to nonlocal
theories by playing with noncommutative actions are conducted without paying
any attention to macro-causality requirements, one cannot help of thinking of
d\'{e}ja vu of what Swieca encountered in the 70s. Even if the motivation has
somewhat changed, the conceptual naivet\'{e} remained on the level of the
Lee-Wick proposal. This is not surprising since the locality issue is one of
the hardest in particle physics, and it seems that this lesson which was
learned the hard way in 60-80, has been partially forgotten and history is
repeating itself.

\section{Some personal recollections and joint projects}

I met Jorge Andr\'{e} Swieca for the first time 1963 in the union hall of the
University of Illinois in Champaign-Urbana when he was on a stop-over coming
from Munich on his return to Sao Paulo, where he was going to defend his
thesis which he finished in Munich under the guidance of his adviser W.
Guettinger\footnote{W.Guettinger was at the ITP in Sao Paulo during the 50's
and Swieca wrote his masters thesis under his guidance and followed him
subsequently to Munich in order to write his PhD Guettinger is a mathematical
physicist who used the (at that time rather new) Laurent Schwarz theory of
distributions in physical problems. His research interests at that time were
very similar to those of Giambiagi to whom Andre Swieca also had a very close
relation. There was also a close cooperation between French mathematics at the
USP in Sao Paulo which led to several visits of Laurant Schwartz.} who also
arranged his visit of the Max Planck institute for physics and astrophysics.
The main purpose of his stop-over at the University of Illinois was to present
himself to Rudolf Haag in order to inquire about the possibility of taking up
a post doc position in Haag's group. He started his work at the University of
Illinois around 1963 and stayed within his group for 3 years.

I met him again when I visited Champaign-Urbana around 1965 for a seminar
talk; at that time he invited me to spend some time in Sao Paulo after his
return. It was only in 1968 that I found the time to spend a couple of months
at the USP in Sao Paulo.

The active members of the Brazilian physics community soon recognized his
extraordinary talent. Without their support he would not have received the
Brazilian Santista science prize already in the late 60s, shortly after his
return from the US. It was given to him for his contributions to the improved
understanding of symmetries and their spontaneous breaking which also has been
the main topic of this essay. Before I continue to write about some episodes
connected with my visits to Brazil, some memories about my first scientic
encounters with Swieca are in order.

During my affiliation with the University of Pittsburgh in the 70's, I felt
attracted by some peculiarities of conformal theories as e.g. problem of how
the Huygens principle of free massless classical fields in even spacetime
dimensions passes to the quantum case. Conformal QFT enjoyed already some
short-lived interests a decade before, but as a result of problems to
reconcile conformal interactions with the particle structure it naturally fell
into disgrace at a time when all attention in QFT was directed towards
dispersion relations and scattering theory.

The starting observation was that some of the zero mass models which were new
at that time, as e.g. the massless Thirring model, did not fulfill Huygens
principle \cite{HSS}, even though by the standards of checking the
infinitesimal form of invariance (commutations with the would be generators)
they were conformally covariant. Instead of a propagation on the mantle of the
light cone, these models propagated inside the cone, which, in analogy to
acoustics, was termed "reverberation". In the setting of Minkowski spacetime
the global propagation even violated causality because timelike distances
inside the light-cone can be transformed into spacelike separations. In order
to have a mathematically solid starting point, Swieca together with V\"{o}lkel
re-visited the zero mass free fields case in order to prove that not only the
Poincar\'{e} generators, but also the remaining conformal generators have a
well-defined mathematical functional analytic definition. The details were
actually quite tricky \cite{SV}. This work was later taken up by Hislop and
Longo \cite{HL} who placed this into the more general context within the
setting of algebraic QFT.

On a second visit to Brazil I started to collaborate with Swieca on structural
properties of interacting QFTs. We understood that anomalous dimensions always
activate the covering of the conformal group as well as the covering of the
(Dirac-Weyl) compactified Minkowski spacetime. This is one of the few cases
where the presence of interactions is directly linked to a change from group
representation to that of its covering\footnote{The idea that the dynamical
aspects of massive QFT could be governed by group representation theory of
noncompact groups was very popular, but these attempts ended in No Go theorems
connected with the name O'Raifeartaigh and Coleman-Mandula.}.

One consequence of the presence of a nontrivially represented diagonolizable
center $\mathbb{Z}$ of the conformal covering (which is in the center of the
field algebra) was that fields which one expected to carry an irreducible
representation of the conformal group in fact only appeared to behave
irreducibly under infinitesimal transformations and therefore admitted a
decomposition with respect to the center of the covering group. The result of
our collaboration was a very rich conformal decomposition theory into simpler
conformal blocks \cite{S-S}\cite{SSV} whose application to the problem of
commutation relations led us into what we called the conformal nonlocal
decomposition theory.

In contradistinction to the undecomposed fields, the component fields seem to
have a simpler timelike commutation structure. Since no interacting
controllable 4-dimensional model existed, we adapted our decomposition theory
to two dimensions. In that case the conformal group factorizes together with
the QFT into two chiral components and our chiral test model was the
exponential of the free massless boson (whose rich charge structure was
already known). These chiral models live on a light ray so that space- and
time- like coalesce to lightlike and the distinction between spacelike
distances and the Huygens region is lost. The commutation relations of the
$Z$-reduced field is that of "anyons" i.e. abelian representations of the
braid group which appear as numerical factors if one changes the order in the
product of two operators. The decomposition theory for the massless Thirring
model is completely analogous.

This nourished the hope that the components of conformal anomalous dimension
fields in higher spacetime dimension have simple commutation relation in the
timelike Huygens region in analogy to chiral models and this may be an
algebraic structure which, if coupled with the spacelike (anti)commutation,
may provide the additional algebraic structure which is necessary for a
classification and construction of higher-dimensional conformal QFT in analogy
to the lightlike plektonic commutation structure of chiral models (where
space- and timelike coalesce to lightlike distances).

Although there have been some exciting new results about the structure of
observable algebras \cite{RT1} \cite{RT2} which by definition live on the
Dirac-Weyl compactified Minkowski spacetime and do not require the
introduction of its covering, the full understanding of higher dimensional
conformal field theory still remains a challenging theoretical problem to date.

The operator-based algebraic research about the global conformal decomposition
theory 1974/75 by Swieca and collaborators came to an halt after it was noted
that the component fields (nowadays called "conformal blocks"). as a result of
their dependence on the central (source and range) projectors associated to
the conformal covering, they were neither ordinary (Wightman)
fields\footnote{Certain components which appear in the decomposition
annihilate the vacuum i.e. they violate the Reeh-Schlieder property ("the
state-field relation") which does not happen for Wightman fields.} nor did
they have a natural euclidean setting and hence they did not fit the
prejudices of those times. Not having been able to fully liberate ourselves
from these prejudices we did not believe that besides our simple illustrations
of a new kind of commutation relations (exotic statistics, abelian braid group
structure) there could be less simple realizations of our general
decomposition theory than those in terms of our exponential Bose fields; for
this reason we directed our attention to other topics.

Several years later Gerhard Mack showed some results he obtained with Martin
Luescher in which they constructed the conformal limit of the Ising QFT from a
representation theory of the conformal energy-stress tensor (the beginning of
the c-quantization, with $c=\frac{1}{2}$). They never published their
interesting results and I sincerely hope that they had other reasons for their
timid attitude than my less than enthusiastic reaction. The end of this story
is well known and needs no detailed presentation: the issue of nontrivial
chiral conformal QFT exploded after Belavin, Polyakov and Zamolodchikov
discovered the first nontrivial family of "minimal" representations of the
energy-stress algebra. This was shortly after Swieca's premature deathe in
December 1980. BPZ, as well as others, also added the new (for physicist)
methodological tool of Kac-Moody algebras and Verma-modules.

It was not difficult to see that our central decomposition theory nicely
harmonizes with the BPZ conformal block decomposition. One could also see that
their commutation relations still represented the braid group, but now some of
the new representations were not abelian (anyons) but rather nonabelian
(plektons) representations of the braid group of the kind as they appeared
naturally in Vaughn Jones mathematical subfactor theory. To understand the
relation between the old work from the time with Swieca and the new BPZ
setting was not a simple matter\footnote{This is not surprizing since one
important mathematical tool namely the representation theory of Kac-Moody
algebras and loop groups did not yet exist or was not known outside
mathematics.}, K.-H. Rehren and myself worked almost two years on this task of
linking the old view about conformal decomposition theory with the new one
\cite{Artin}.

In the early 70s when the grip of the military dictatorship on public
institutions especially on universities was getting tighter, many theoretical
physicists, including Andr\'{e} felt more secure at the Ponteficia
Universidade Catolica (PUC) in Rio de Janeiro, a non state run university
under the umbrella of the (at that time very progressive) catholic church. A
bypass heart surgery forced him to follow medical advice and look for a
quieter place in the countryside. He continued his research at the smaller
Federal University in Sao Carlos, only to realize some time after that the
advice was not so good after all. Whereas at the PUC in Rio he was surrounded
by well-intentioned and supportive colleagues, in Sao Carlos he had to engage
in exhaustive struggles with the department chairman in order to salvage some
agreements and promises which were made to him before coming. This aggravated
his health and certainly contributed to his premature death at the end of 1980.

Different from the Pinochet regime in Chile, the US was probably not directly
involved in the installation of its Brazilian counterpart, but the Brazilian
generals received US sympathy and support after their military take over in a
coup. At that time there was a deep gap between the proclaimed US democratic
ideals and the consequences of their realpolitik in the name of
anti-communism. But apart from my short visits to Brazil, these problems were
removed from my life as a professor of physics in Pittsburgh; in any case I
enjoyed my 8 year stay in the US, and apart from a critical distance to
certain political developments, it was my impression that Jorge Andr\'{e} felt
the same way when he spent 2 years in the US; although we rarely discussed
politics. Only some years ago I learned that around 1970 the military regime
in Brazil offered him a diplomatic post in Israel (presumably that of a
scientific attache) which he declined, certainly because he found the idea to
represent a dictatorship not appealing.

With the shared scientific background as a result of having been a member of
the "Haag school" of QFT\footnote{Rudolf Haag is the protagonist of the
algebraic approach to QFT an approach which tries to avoid the quantization
paralellism to classical field theories in favor of a more intrinsic
understanding.}, it was quite easy to agree with him on what are the
interesting particle physics problems and to use our common stock of
conceptual and mathematical knowledge to try to solve them. My first trip to
Brazil in 1968 was the beginning of many more visits to the USP in S\~{a}o
Paulo and later to the PUC in Rio de Janeiro and the UF in Sao Carlos.

As mentioned in the previous section, in the first half of the 70 there was a
flurry about some quasiclassical observations on certain two-dimensional QFTs
in which the quasiclassical particle spectrum seemed to be exact \cite{DHN}.
This signalled some form of integrability; but contrary to the integrability
in QM (e.g. the hydrogen atom), the field theoretic setting required some new
ideas. Concentrating on a particular model, it was not difficult to see that
the quasiclassical spectrum originates from a simple 2-particle scattering
matrix together with factorization property and a fusion picture for higher
bound-states from the lowest one. This was a resurrection of the old S-matrix
bootstrap picture but now in the more limited context of two-dimensional
factorizing S-matrices \cite{STW}.

Within a short time a group of enthusiastic young members of the newly formed
QFT group at the Free University of Berlin around Michael Karowski and Peter
Weisz (who I had the pleasure to advise) found the general solution: the
ingredients of the old (and already in the 70s abandoned) S-matrix bootstrap
approach in two-dimensional theories with only elastic scattering, if
augmented with factorization and a fusion mechanism for bound states
consistent with the nuclear democracy principle, worked in a beautiful manner.

Upon taking notice of these developments, Jorge Andre got quite excited about
these results. He recognized the potential of these tfactorizing models as
theoretical laboratories for testing all kinds of field theoretic ideas
outside the perturbative setting. His previous experience with simpler
two-dimensional models as the before mentioned zero mass exponential bosons,
the closely related Thirring model, and a field theoretic version of
Kadanoff's results on order/disorder variables\footnote{The topics led to
several master- and PhD thesis by his students e.g. \cite{Ma}.} greatly
facilitated his start. An article which reflects the state of art can be found
under \cite{Fort}.

Within his short time he made important contributions and introduced a whole
new generation of Brazilian physics students to these new problems. In this
way he played a crucial role in the formation of the first generation of
particle theorist in Brazil. I had the good luck to enter particle physics in
interesting times and to meet and collaborate with remarkable individuals as
Jorge Andr\'{e} Swieca.

Most physicists in Brazil and even many people outside physics \ knows the
name Swieca; this is partially due to the fact that an important yearly
occuring physics summer school organized by the university of Sao Paulo is
called \textit{Jorge Andr\'{e} Swieca Summer School in Particle and Fields}.
This honor is more than justified by the fact that the particle theory
research in Brazil started with Swieca and his school. But few physicists of
the younger generations are familiar with the actual content of Swieca's
contributions to particle physics and the legacy of his work in present
developments. Some of the problems he proposed, investigated and, in some
cases, completely solved by him led to questions which are still in the
forefront of discussions. They intertwine the present research in QFT in an
interesting way with the particle theory of the 60/70; hence a fresh look at
Swieca's work as attempted in the present essay is more than just doing
scientific archeology.

Jorge Andr\'{e} Swieca was not Brazilian by birth. He was born 1936 in Warsaw
and fled with his parents from the Russian occupied part of Poland shortly
after it was divided between Hitler and Stalin. The odyssey, which started
with a trip on the Transsib railroad to Wladivostok (with the 3 year old
Andr\'{e} without papers hidden underneath a seat at each control) and then
continued with the ferry to Yokohama, only ended after spending two years in
Japan and Argentine before his parents finally were able to settle down in Rio
de Janeiro/Brazil. Although the Nazi persecution only affected him through the
flight of his family, he later on (for example during his stay at the MPI in
Munich) met a physicist who had a number tattooed in his arm.

As many Jewish survivors of the holocaust, Jorge Andr\'{e} had a trauma which
can be summerized by the agonizing thought "why I and not the others" which
(in my opinion) contributed at least as much to his taking his life as the
detoriation of his physical state after his bypass heart surgery.

\end{document}